\documentclass[twocolumn,trackchanges]{aastex631}
\newcommand{\teff}{T_{\mathrm{eff}}}

\usepackage{amsmath}
\usepackage{bm}
\usepackage{amsthm,amsmath,amssymb}
\usepackage{mathrsfs}

\begin{document}

\title{Spectroscopic ages for 4 million main-sequence dwarf stars from LAMOST DR10 estimated with data-driven approach}

\author{Jia-Hui Wang}
\affiliation{National Astronomical Observatories, Chinese Academy of Sciences, Beijing 100012, China}
\affiliation{School of Astronomy and Space Science, University of Chinese Academy of Sciences, Beijing 100049, China}
\email{wangjh@bao.ac.cn}

\author{Maosheng Xiang}
\affiliation{National Astronomical Observatories, Chinese Academy of Sciences, Beijing 100012, China}
\affiliation{Institute for Frontiers in Astronomy and Astrophysics, Beijing Normal University, Beijing, 102206, China}
\email{msxiang@nao.cas.cn}

\author{Meng Zhang}
\affiliation{National Astronomical Observatories, Chinese Academy of Sciences, Beijing 100012, China}

\author{Ji-Wei Xie}
\affiliation{School of Astronomy and Space Science, Nanjing University, Nanjing 210023, China}
\affiliation{Key Laboratory of Modern Astronomy and Astrophysics, Ministry of Education, Nanjing 210023, China}

\author{Jian Ge}
\affiliation{Shanghai Astronomical Observatory, Chinese Academy of Sciences, Shanghai 200030, China }

\author{Jinghua Zhang}
\affiliation{South-Western Institute for Astronomy Research, Yunnan University, Chenggong District, Kunming 650500, China }
\affiliation{CAS Key Laboratory of Optical Astronomy, National Astronomical Observatories, Chinese Academy of Sciences, Beijing 100101, China}

\author{Lanya Mou}
\affiliation{National Astronomical Observatories, Chinese Academy of Sciences, Beijing 100012, China}
\affiliation{School of Astronomy and Space Science, University of Chinese Academy of Sciences, Beijing 100049, China}

\author{Ji-Feng Liu}
\affiliation{National Astronomical Observatories, Chinese Academy of Sciences, Beijing 100012, China}
\affiliation{School of Astronomy and Space Science, University of Chinese Academy of Sciences, Beijing 100049, China}
\affiliation{Institute for Frontiers in Astronomy and Astrophysics, Beijing Normal University, Beijing, 102206, China}

%\email{jfliu@nao.cas.cn}

% \collaboration{20}{(AAS Journals Data Editors)}

% \author{F.X Timmes}
% \affiliation{Arizona State University}
% \affiliation{AAS Journals Associate Editor-in-Chief}

% \author{Amy Hendrickson}
% \altaffiliation{AASTeX v6+ programmer}
% \affiliation{TeXnology Inc.}

% \author{Julie Steffen}
% \affiliation{AAS Director of Publishing}
% \affiliation{American Astronomical Society \\
% 1667 K Street NW, Suite 800 \\
% Washington, DC 20006, USA}



\begin{abstract}
Stellar age determination for large samples of stars opens new avenues for a broad range of astronomical sciences. While precise stellar ages for evolved stars have been derived from large ground- and space-based stellar surveys, reliable age determination for cool main-sequence dwarf stars remains a challenge. In this work, we set out to estimate the age of dwarf stars from the LAMOST spectra with a data-driven approach. We build a training set by using wide binaries that the primary component has reliable isochrone age estimate thus gives the age of the secondary. This training set is further supplemented with field stars and cluster stars whose ages are known. We then train a data-driven model for inferring age from their spectra with the XGBoost algorithm. Given a spectral signal-to-noise ratio greater than 50, the age estimation precise to 10\% to 25\% for K-type stars, as younger stars have larger relative errors. Validations suggest that the underlying information used for our age estimation is largely attributed to the LAMOST spectral features of chemical abundances. It means our result is a manifestation of stellar chemical clock effectively acted on LAMOST spectra ($R\simeq1800$). Applying our model to the LAMOST DR10 yields a massive age catalog for $\sim4$ million dwarf stars. Statistical properties, such as the age distribution, age-abundance and age-stellar activity relations of the sample stars are discussed. The catalog is publicly accessible and can be helpful for extensive sciences from detection and characterization of Earth-like planets to Galactic archaeology.
\end{abstract}



\section{Introduction} \label{sec:intro}

Stellar ages play a crucial role in understanding the evolution of stars, the formation history of galaxies, and in the search for exoplanets and understanding their evolution \citep{Soderblom_2010,Xiang_2022,ge2022et, ChenD2023}. Accurate and precise determination of stellar ages is thus of great importance but usually a challenging task that relies on indirect estimation under the deep involvement of stellar evolution models \citep{Soderblom_2010}.

During the past decade, two age dating approaches, isochrone fit and asteroseismology, both based on stellar evolutionary models, have achieved remarkable progress in determining the ages of large samples of field stars. The first approach fits the measured stellar atmospheric parameters with predictions of stellar evolutionary models. This approach is especially effective for stars in the main-sequence turn-off (MSTO) and subgiant phases, during which their positions on the $T_{\rm eff}$-$\log$~$g$ (Kiel) or temperature-luminosity (Hertzsprung-Russell; HR) diagram are most sensitive to their ages. Owing to the availability of precise stellar atmospheric parameters for millions of stars from large spectroscopic surveys, robust ages for millions of MSTO and subgiant stars have been derived \citep{2015RAA....15.1209X, 2017ApJS..232....2X}. Combing the spectroscopic parameters with precise luminosity from astrometric parallax of the Gaia mission \citep{2016A&A...595A...1G,2018ApJ...866...39B,2022ApJ...933...94B}, extensive works have been carried out to estimate the ages of stars across a broad range in the parameter space with isochrone fit \citep[e.g.][]{2018MNRAS.481.4093S, 2019MNRAS.489.1742F, 2019A&A...629A.127M, 2021MNRAS.506..150B, 2021ApJ...922..189V, Xiang_2022, 2022A&A...658A..91A, 2023MNRAS.523.1199S, 2023RAA....23b5020W}. In particular, it has been suggested that, with such a method, stellar age determination can reach a high precision of only a few percent relative errors for subgiant stars \citep{Xiang_2022, 2024arXiv240718307N}. 

Asteroseismic analysis of high-precision light curves can precisely yield the size and mass of stars, thus providing crucial constraints of age for evolved stars. A combination of oscillation frequency delivered from the Kepler mission \citep{2010AAS...21510101B} with spectroscopic stellar atmospheric parameters has enabled robust asteroseismic age determination for thousands of evolved stars \citep[e.g.][]{2017RAA....17....5W, 2018MNRAS.475.3633W, 2023MNRAS.520.1913W, 2017ApJS..233...23S, 2018MNRAS.475.5487S, 2021A&A...645A..85M, 2022ApJ...927..167L}. 

On the other hand, the age determination for dwarf stars, which dominate the stellar populations, remains a great challenge. Asteroseismic parameters has been attempted to constrain the ages for main-sequence dwarf stars \citep{2017ApJ...835..173S}, but it works mainly for stars with relatively warm temperatures ($T_{\rm eff}\gtrsim5400$~K), whereas it is much harder for cooler stars as isochrones of different ages are almost indiscernible based on their mass and atmospheric parameters. Gyrochronology, which estimates the stellar age based on a color-dependent relation between age and rotation period, has been proposed as an empirical way of age dating for dwarf stars \citep{ 2003ApJ...586..464B,2007ApJ...669.1167B,2008ApJ...687.1264M,2015MNRAS.450.1787A, 2022arXiv221204515L,2024ApJS..271...19Y}. Similarly, several other empirical approaches, such as age-stellar activity relation and age-lithium abundance relation have been also proposed as a practical age clock for main-sequence dwarf stars \citep[e.g.,][]{1972ApJ...171..565S, 1995ApJ...438..269B, 2005A&A...442..615S}. For these approaches, a good calibration of the empirical age relations is essential but still far from perfect due to either observational challenges or intrinsic complications. As a consequence, the availability of these approaches is mostly restricted to stars younger than the Sun \citep{1995ApJ...438..269B, 2005A&A...442..615S,  2015Natur.517..589M,2015EPJWC.10105006V,2021MNRAS.506L..50T}. 

Stellar chemical abundances have also been suggested to be good age indicators. First, observations show clear correlation among age, [Fe/H], and $[\mathrm{\alpha} / \mathrm{Fe}]$ \citep[e.g.][]{Haywood_2013, Xiang_2017}, as the metal-poor, high-alpha stars are generally old, while the metal-rich young stars are relatively young. Second, the abundance ratios that involve a neutron-capture element such as [Y/Mg] has been widely used as chemical clock \citep[e.g.][]{2019A&A...631A.171S, 2022ApJ...936..100B, 2022MNRAS.517.5325H}. Similar to other empirical age relations as mentioned above, a good calibration is essential for age estimation with the chemical clock. This especially matters considering that age-abundance relation may vary with location across the Galaxy \citep{2024MNRAS.528.3464R}. Currently, the application of chemical clocks are mainly restricted to stellar abundances from high-resolution spectroscopy, and the sample stars are thus mostly located at solar neighbourhood. 

%Data-driven approach has opened up an avenue for stellar parameter estimation in the era of large sky surveys \citep{2015ApJ...808...16N, 2017ApJ...849L...9T, 2019ApJS..245...34X, 2019MNRAS.490.4740B, 2020ApJS..246....9Z}. 
Alternatively, data-driven approach for stellar age estimation has been applied to large survey data sets, and delivered ages for millions of red giant stars \citep[e.g.][]{2015MNRAS.451.2230M, 2016ApJ...823..114N, 2017ApJ...841...40H, 2019MNRAS.484.5315W, 2020ApJS..249...29H, 2021ApJ...922..145Z}. This is mainly attributed to the availability of a good set of age reference stars, in particularly the asteroseismic age sample as mentioned above. 

In this work, we set out to derive the age of dwarf stars from the LAMOST spectra with a data-driven approach. The LAMOST spectra cover the full range of optical wavelength window \citep{2012RAA....12.1197C, 2022Innov...300224Y}, such that in principle they contain multiple pieces of information that can indicate the age of a star, for instance, the stellar atmospheric parameters and the chemical abundances. Compared to that for MSTO and subgiant stars, the stellar atmospheric parameters for main-sequence dwarf stars show much less sensitivity on their ages. Therefore, age estimation method suitable for the formers such as isochrone fitting may fail to yield realistic results for the latter, resulting age estimates with both poor precision and large deviation. A data-driven approach is thus an optimized choice to derive ages from the LAMOST spectra for dwarf stars.   

The foundation for a data-driven approach is a benchmark (training) sample of stars with precise age determinations. Wide binaries (WB) provide a good solution for this purpose. Millions of wide binaries have been identified from the Gaia astrometric survey \citep{2018MNRAS.480.4884E, 2021MNRAS.506.2269E}. The component stars in a wide binary system have identical ages but evolve independently, so that for a binary system composed of a dwarf and a subgiant star, the dwarf's age can be known from the latter. Unlike other tracers that may be biased in parameter space (e.g., age, metallicity, HR diagram), such as open and globular clusters, wide binaries extensively distribute in a broad range of parameter space, making them a good training set for data-driven age estimation for large surveys like LAMOST. 

We adopt wide binaries, accomplished with some cluster member stars and field stars that have LAMOST spectra and known ages as the training set, and use the {\sc XGBoost} algorithm to build the data-driven model for the age estimation. The paper is organized as follows. Section \ref{sec:data} describes our training samples. Section \ref{sec:method} introduces our data-driven model. Section \ref{sec:result} presents the resultant stellar age catalog and validations. Section \ref{sec:discussion} presents a discussion of the results. Section \ref{sec:conclusion} concludes.

\section{Data} \label{sec:data}
%\subsection{Training samples}
\subsection{The LAMOST DR10 spectra}
The tenth data release (DR10) of the LAMOST Galactic surveys \citep{2012RAA....12..723Z, 2012RAA....12..735D, 2014IAUS..298..310L, 2022Innov...300224Y} released 11,817,430 low-resolution ($R\simeq1800$) spectra, covering the full optical range of 3700--9000\AA. The spectroscopic targets are uniformly drawn from the photometric color-magnitude diagram (CMD) with simple yet non-trivial selection function \citep{2015ApJ...799..135Y, 2017RAA....17...96L,2018MNRAS.476.3278C}, leading to a massive spectroscopic sample covering the full stellar parameter space. While the majority of the spectroscopic stars are main-sequence dwarfs, as they are the dominant population of our Galaxy, a large number of stars in brief evolution phases, such as the main-sequence turn-off and subgiant phases, are also effectively sampled \citep[e.g.][]{Xiang_2017, Xiang_2022}.

The allocation of 4000 LAMOST fibers is implemented with 4000 robots, each having an allocation area of 2 arc minutes. As a consequence, only one star can be allocated within a 2-arcmin area for a single visit to avoid fiber cross-talk. However, because the survey has multiple (usually 2 to 3) visits and brightness categories of plates (very bright, bright, medium bright, faint) for a given field defined by a central bright star, as well as overlaps of the adjacent fields, stars can be sampled with much higher spatial density \citep{2014ApJ...790..110L,2015ApJ...799..135Y}. This is important for our purpose as we define training sample using wide binary systems that both of the components are observed by LAMOST.  

In this work, we use normalized spectra for the age estimation. Our normalization scheme is the same as \citet{2019ApJS..245...34X}. Specifically, the LAMOST spectra $f(\lambda)$ are divided by a smoothed version of itself $\bar{f}(\lambda)$ by convolving with a Gaussian kernel, 
\begin{equation}
      \bar{f}(\lambda_i) = \frac{\Sigma_jf(\lambda_j)\omega_j(\lambda_i)}{\Sigma_j\omega_j(\lambda_i)},
    \end{equation}
    \begin{equation}
      \omega_j(\lambda_i) = \frac{1}{\sqrt{\pi}L}\exp\left(-\frac{(\lambda_j-\lambda_i)^2}{L^2}\right),
    \end{equation}
where the Gaussian kernel width $L$ is adopted as a constant value of 50{\AA}. The normalized spectrum $f_{\rm n}(\lambda)$ is then obtained via
 \begin{equation}
     f_{\rm n}(\lambda) = \frac{f(\lambda)}{\bar{f}(\lambda)}.
 \end{equation}

\subsection{The age training sample}
To obtain a good training sample is the key for data-driven age estimation. Our training sample is constructed combining four data sets, 
\begin{itemize}
    \item First, we build a sample of about 1200 wide binaries for which both components have a LAMOST spectrum, and the primary stars have an effective temperature higher than 5400~K to allow reliable isochrone age estimation.  
    \item Second, to increase the number of old stars in the training sample, we supplement the wide binary sample with about 900 stars that have $T_{\rm eff}>5400$~K and an isochrone age older than 6~Gyr, with age uncertainty less than 10\%. 
    \item Third, we further supplement the training sample with 400 high-$\alpha$, old stars that have $T_{\rm eff}<5400$~K and age older than 9~Gyr. Their ages are inferred from the tight age-metallicity relation of high-$\alpha$ stars. 
    \item Finally, we add the member stars from a number of open clusters to the training set. 
\end{itemize}

Approximately 30\% of the spectra in LAMOST DR10 are repeat observations. In our training sample, for stars with repeat observations, we only use the spectrum with the highest signal-to-noise ratio ($S/N$). 

\subsubsection{The wide binary sample} \label{subsec:tables}
We cross-match LAMOST and the Gaia wide binary catalog of \citet{2021MNRAS.506.2269E} using a 3-arcsec criterion, approximately the aperture size of LAMOST fibers. For the Gaia wide binary catalog, we only adopt those with higher than 90\% confidence of being true binary, while discarding the others to avoid chance alignment. We restrict our training sample to binaries that both components have LAMOST spectra with $S/N$ higher than 20 in the SDSS $g$ band. We further require all the stars to have stellar parameters and abundances estimates from the LAMOST DR9 abundance catalog (Zhang et al. in prep.), which is an updated version of the LAMOST DR5 abundance catalog derived with the {\sc DD-Payne} \citep{2019ApJS..245...34X}. These criteria lead to 2893 pairs of binaries. 

% To obtain a clean wide binary sample, we further examine their radial velocity (RV) and metallicity ([Fe/H]) from the LAMOST spectra.
\citet{2021MNRAS.506.2269E} selected binary candidates based on their positions, parallaxes, and proper motions from Gaia eDR3. To obtain a cleaner wide binary sample, we further examine their radial velocity (RV) and metallicity ([Fe/H]) from the LAMOST spectra, which were not considered in El-Badry’s work. Stars in a wide binary system are expected to share similar radial velocities, with an RV difference no larger than the maximal relative motion of the components w.r.t. the mass center, which is a few km/s in typical given the wide orbit. Also, the component stars are expected to have almost identical metallicity as they are siblings born from the same gas cloud.

\begin{figure*}[ht!]
\centering
\includegraphics[width=1\linewidth]{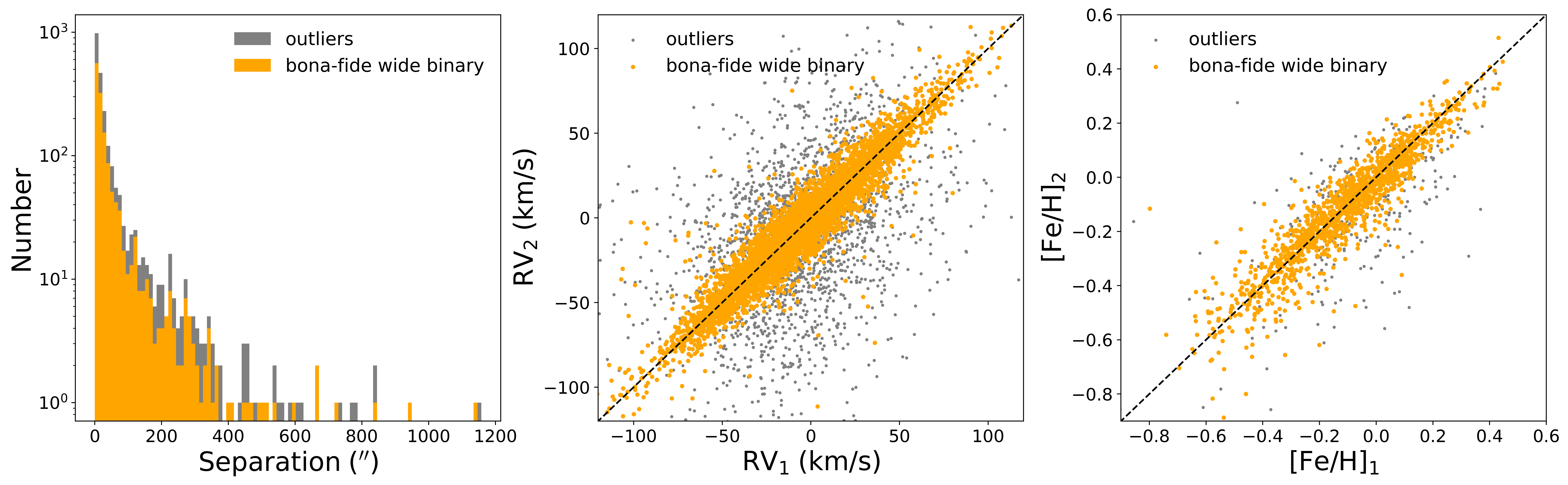}
% \plotone{binary.png}
\caption{The wide binary sample. The left panel shows the distribution of separation angle of the wide binaries, with a lower separation angle cut of 3". The middle and right panel show the comparisons of radial velocity and metallicity, respectively, between the individual component stars of the wide binary systems. The orange color delineates bona-fide binaries selected from the radial velocity and metallicity criteria (see text). The gray color delineates outliers, which are probably recognized as wide binaries because of chance alignment. 
\label{fig:binarypara}}
\end{figure*}

Figure \ref{fig:binarypara} presents a comparison of the RV and [Fe/H] between the binary components. To identify outliers from the comparison, we first define the RV and [Fe/H] criteria as in Eq.\ref{eq1} and Eq.\ref{eq2}: 

\begin{equation}\label{eq1}
\mathrm{\Delta}_{\text {RV }}=\frac{\left(\mathrm{RV}_1-\mathrm{RV}_2\right)}{\sqrt{\sigma_{\mathrm{RV}_1}^2+\sigma_{\mathrm{RV}_2}^2+
\sigma_{\rm intrinsic}^{2}}
},
\end{equation}

\begin{equation}\label{eq2}
\mathrm{\Delta}_{[\mathrm{Fe} / \mathrm{H}]}=\frac{\left([\mathrm{Fe} / \mathrm{H}]_1-[\mathrm{Fe} / \mathrm{H}]_2\right)}{\sqrt{\sigma_{[\mathrm{Fe} / \mathrm{H}]_1}^2+\sigma_{[\mathrm{Fe} / \mathrm{H}]_2}^2+
\sigma_{\rm intrinsic}^{2}}
},
\end{equation}
where RV$_1$ (${\rm [Fe/H]_1}$) refers to the radial velocity (metallicity) of the primary star that has bright Gaia G magnitude, and $\sigma$ refers to the measurement uncertainty. $\sigma_{\rm intrinsic}$ refers to the intrinsic differences in binaries. In the processing, we assume $\sigma_{\rm intrinsic}$ is 0, which will lead us to a more stringent judgment of true binaries, but give us a purer sample. Binaries that either the RV or the [Fe/H] between the components deviate by larger than 3$\sigma$ are treated as chance alignment and discarded from the sample, except for cool dwarfs with $T_{\rm eff} < 4800$~K and $\log~g > 4$. For these cool dwarfs, only the RV criterion is applied, as their [Fe/H] estimate may suffer large systematics \citep[e.g.][]{2023ApJ...950..104N}.

\begin{figure*}[ht!]
\centering
\includegraphics[width=0.95\linewidth,]{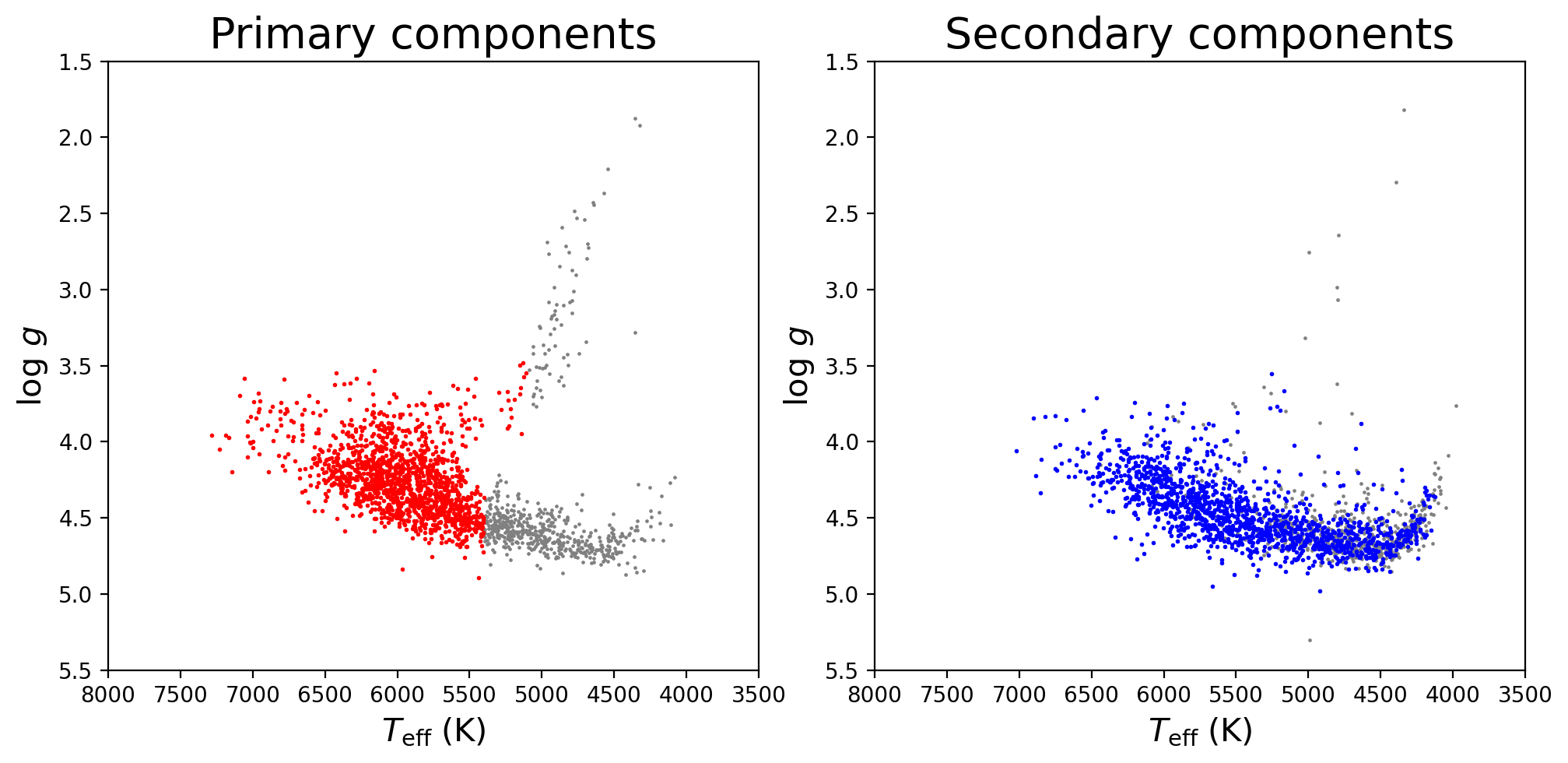}
\caption{Distribution of wide binaries in the $T_{\rm eff}-\log~g$ diagram. The left and right panels show respectively the primary and secondary components, defined in \citep{2021MNRAS.506.2269E}. The colored dots are reference sample stars for age estimation in the current work. For these binaries, the primaries are in either subgiant of main-sequence (turn-off) evolutionary phase and their ages can be reliably determined with isochorne fitting, while the secondaries have a broad distribution in temperature, and are adopted as the training set of our data-driven age estimation.  
%The figure shows the selected binaries. The left panel shows the primary stars, with the MSTO stars in dark red and the subgiants in orange. The right panel shows the secondary stars, which have temperatures ranging from 4000 K to 8000 K and are dominated by the main sequence stars.
\label{fig:binarytl}}
\end{figure*}

Furthermore, only the binary systems in which the primary has $T_{\rm eff}>5400 \rm K$, which allow robust isochrone age estimation from their atmospheric parameters, are adopted as our training sample. For subgiant stars specifically, we adopt $T_{\rm eff}>5100 \rm K$ instead of $T_{\rm eff}>5400 \rm K$ (Figure \ref{fig:binarytl}). 

Ultimately, these criteria leads to 1213 pairs of wide binaries. The right panel of Figure \ref{fig:binarytl} illustrates that the training sample covers a broad range of the HR diagram, and has especially a good sampling for the cool dwarf stars. 

We determine the age of the primary star components by fitting their atmospheric parameters to the MESA Isochrones \& Stellar Tracks \citep[MIST;][]{2016ApJ...823..102C} with a Bayesian approach, utilizing the DD-Payne spectroscopic parameters ($T_{\rm eff}$, $\rm M_{K, spec}$, [M/H]) of LAMOST DR9, multi-band photometric magnitudes of the Gaia \cite{2021A&A...649A...3R, 2021A&A...649A...1G} and 2MASS \citep{2006AJ....131.1163S}, as well as the Gaia astrometric eDR3 parallax \citep{2021A&A...649A...1G} as observables. Here the overall metallicity [M/H] is converted from the DD-Payne [Fe/H] and [$\alpha$/Fe], following the method of \citet{1993ApJ...414..580S}. More details about the age determination are referred to \citet{Xiang_2022}.

\subsubsection{The old, warm star sample}
We found that the number of old stars in the binary sample is too small to allow for building a robust data-driven model for age estimation. We therefore supplement the wide binary sample by some field stars that have precise isochrone ages to increase the number of old stars in the training sample. 

\begin{figure}[ht!]
\centering
\includegraphics[width=1\linewidth,]{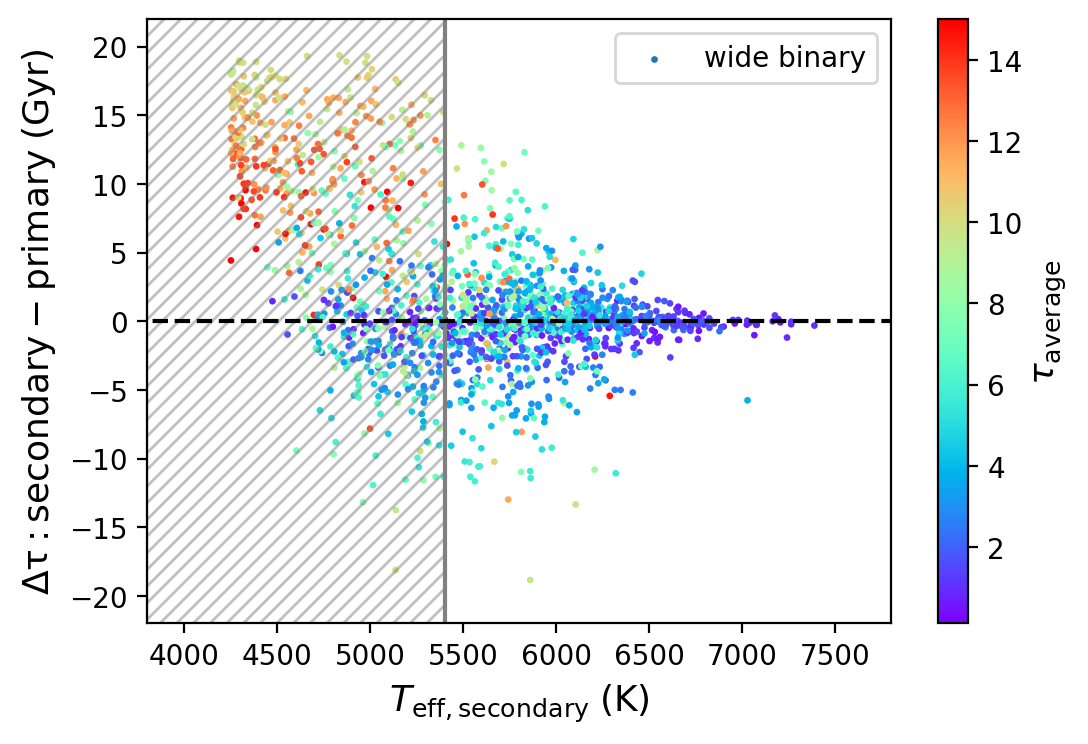}
\caption{The difference in isochrone age estimates between the secondary and primary stars of wide binaries as a function of effective temperature of the secondary. All the primary stars have an effective temperature higher than 5400~K. The color represents the average age of the binaries. The shaded area delineates regime where the isochrone ages are unreliable.}
\label{fig:lower}
\end{figure}

In doing so, we first determine the lower temperature border below which the isochrone age estimate becomes unreliable, utilizing wide binaries as the test beds. We adopt binary systems for which the primary have an effective temperature higher than 5400~K, and investigate the differences in the isochrone age estimates between the secondary and primary as a function of effective temperature of the secondary. As shown in Figure~\ref{fig:lower}, at the high temperature side the ages of the secondary and primary are estimated consistently, with only a small difference. However, the scatter becomes larger as the temperature decreases. Especially, for stars below about 5400~K, a large proportion of them shows large discrepancy, reflecting the fact that isochrone age estimates in regime is hard. 

Therefore, in order to make use of the isochrone ages, we only add main-sequence stars with effective temperature higher than 5400~K as a supplementary to wide binary sample. We uniformly select stars based on their effective temperature and ages, requiring an age estimate older than 6~Gyr, and relative age error smaller than 10\%. Ultimately, the selection leads a sample of 927 old field stars. 

\subsubsection{The old, cool star sample}
For main-sequence stars cooler than 5400~K, the isochrone ages become problematic, and alternative method is required to obtain accurate ages. \citet{Xiang_2022} found that for high-$\alpha$ stars, there exists a very clear and tight age-metallicity relation in the range of $-1.0\lesssim {\rm [Fe/H]} \lesssim0.5$, and $8\lesssim\tau\lesssim13$~Gyr, and they provided a linear fit to the relation as $\bar{\tau}([\mathrm{Fe} / \mathrm{H}])=9.38+(-3.79) \times[\mathrm{Fe} / \mathrm{H}]$. Intrinsic dispersion to this relation is estimated to be only $\sim0.8$~Gyr. 

In order to add cool old stars into the training set, here we randomly select 400 cool ($\teff<5400$~K) stars that belong to the high-$\alpha$ sequence in the [Fe/H]-[$\alpha$/Fe] plane, and estimate their ages using the above age-[Fe/H] relation. We assign a 10\% error in the resultant age estimate. 

\subsubsection{Age training sample from clusters}
In addition to all the samples listed above, we also include member stars of 4 open clusters (OCs) observed by LAMOST in our training sample. OCs are good complementary training samples at the younger end ($\tau \lesssim1$~Gyr). All member stars in a single cluster are assigned to be a constant age, adopted from literature value. Table~\ref{tab} presents a detailed list of the information for the clusters adopted. For cluster stars, we assign 10\% error in their ages.

\begin{figure}[ht!]
\centering
\includegraphics[width=1\linewidth,]{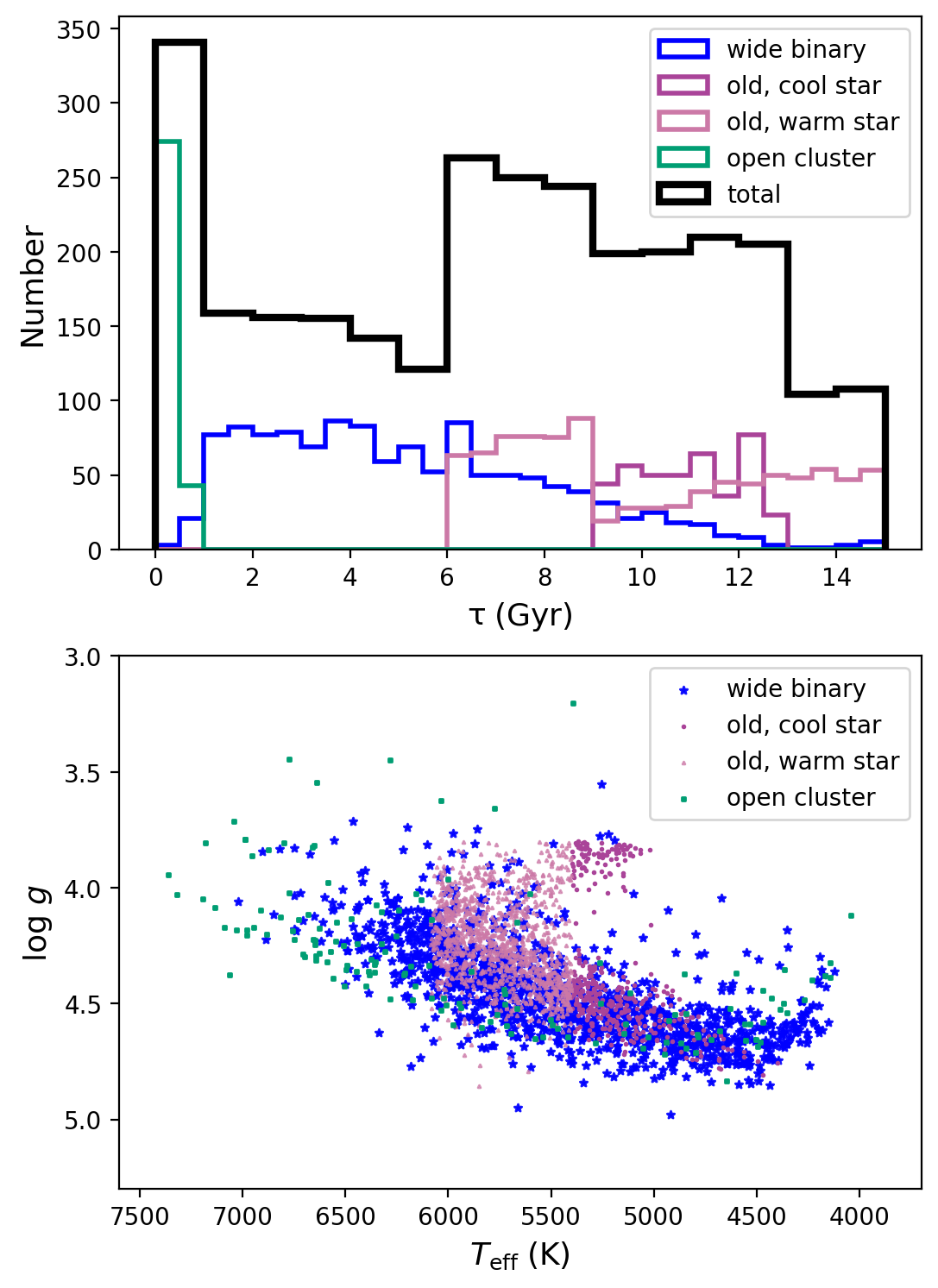}
\caption{Distribution of the training samples. The upper panel shows their age distribution. Different colors represent different types of samples as marked in the figure. Black thick line represent the overall age distribution. The lower panel shows their distribution in the $T_{\rm eff}$-$\log~g$ diagram.}
\label{fig:ad}
\end{figure}

The age distribution of all training samples is shown in Figure~\ref{fig:ad}, with different components marked in different colors.

% Table 1
\begin{deluxetable*}{lccc}

\tabletypesize{\scriptsize}
\tablewidth{0pt} 
\tablecaption{Age of clusters adopted in our training sample  \label{tab}}
\tablehead{
\colhead{Cluster name} & \colhead{Number of stars} & \colhead{Age(Gyr)} & \colhead{Reference}
}
% \colnumbers
\startdata 
Hyades   & 61    & 0.625     & \citet{1998cnm}\\ 
NGC2168  & 154 & 0.175     & \citet{Bouy_2015}\\
NGC2099  & 78   & 0.346-0.55& \citet{Wu_2009,Hartman_2009}\\
Pleiades & 68   & 0.115     & \citet{1996ApJ...458..600B}\\
\enddata

\tablecomments{
1. "Number of stars" depends on the LAMOST observations. \\ 
2. Only half of the member stars are used for training, while the other half is used for validation.
}
\end{deluxetable*}

\section{Method} \label{sec:method}
In this section, we introduce our data-driven model for age determination based on {\sc XGBoost} algorithm, as well as data pre-processing with principal component analysis (PCA), followed by an examination of the algorithms with a validation set.

\subsection{EXtreme Gradient Boosting}

Extreme Gradient Boosting ({\sc XGBoost}) is a powerful machine learning algorithm used for solving classification and regression problems \citep{Chen_2016}. It is an ensemble learning method that boosts the performance of predictions by combining multiple weak learners, typically decision trees, into a strong learner. The core idea behind {\sc XGBoost} is to iteratively train a sequence of decision tree models and adjust the weights of each model based on their prediction errors. {\sc XGBoost} uses an objective function to guide model training and optimizes model performance by minimizing the objective function, which contains the loss function $\mathcal{L}$ as well as the regularizations.

In our model, the spectra after principal component analysis (PCA) are used as inputs to the {\sc XGBoost} model, which is a $2857\times 20$ dimensional array. We construct the regression relationship between spectra and stellar age by {\sc XGBoost}. We customized the loss function $\mathcal{L}$ and considered the age error obtained from the MIST isochrones. Prior to this we mapped age and age error to logarithmic space as shown in Eq.\ref{eq4}.

\begin{equation}\label{eq4}
\begin{aligned}
&\log\tau=\log_{10}\tau \\
&\sigma_{\log\tau}=\sigma_\tau/\tau\log_{10}(e)\\
&\mathcal{L} =\sum_i\left(\frac{{\log\tau}_{\rm {pred, i}}-{\log\tau}_{\rm ref,i}}{\sigma_{\log\tau,i}}\right)^2
\end{aligned}
\end{equation}
where $\tau$ is the reference age of the training sample from the MIST stellar isochrones, $\sigma_\tau$ is the age error of the training sample. ${\log\tau}_{\rm pred}$ is the age predicted by the {\sc XGBoost} model.

In order to avoid overfitting and ensure the convergence of the training process, we fine-tuning the learning rates, number of iterations, and hyperparameters for the model.

%\sout{Considering that the ages of our samples are less distributed at the older end ($\tau \gtrsim10$~Gyr) and the younger end ($\tau \lesssim1$~Gyr), we optimize the fit at both ends by adjusting the weights of the model in stages. The method, on the one hand, takes the unweighted training results as the initial value, which is to preserve the spectral-age relationship learned from the middle-age samples (1~Gyr$\lesssim \tau \lesssim10$~Gyr). On the other hand, in the case of limited and imbalanced samples, we increase the weights at both the older and younger ends in training based on the number of samples in different age bins, which can improve the learned spectrum-age relationship at both ends.} 
%In order to promote the age estimation of low-temperature dwarf stars, 
%In the training process, we specifically set a larger weight to the wide binary sample based their number density in the age bins.

\subsection{Spectral Principal Component Analysis}

Robustness of a machine learning method is usually very sensitive to data noise. For the LAMOST spectra, most of the features that are informative for stellar physical parameters are in the blue wavelength regime($\lesssim5000$~\AA), which only has low $S/N$ for cool stars. Therefore, a de-noising pre-processing of the spectra is helpful to improve the robustness of the age estimates. Here we perform a principal component analysis (PCA) on the spectra for such a purpose. 

PCA converts, in a variance-conservative manner, most of the information of a high-dimensional data set into a small number of principal components via linear combinations of orthogonal basis functions, while leaving the noise information to other components with small variance. Reconstruction of a noise-reduced spectrum with PCA has been applied to the LAMOST spectral analysis in previous work \citep[e.g.][]{2021ApJS..253...22X}. Here we only briefly introduce our model choice, but we omit the math aspect.  

\begin{figure}[ht!]
\centering
\includegraphics[width=1\linewidth,]{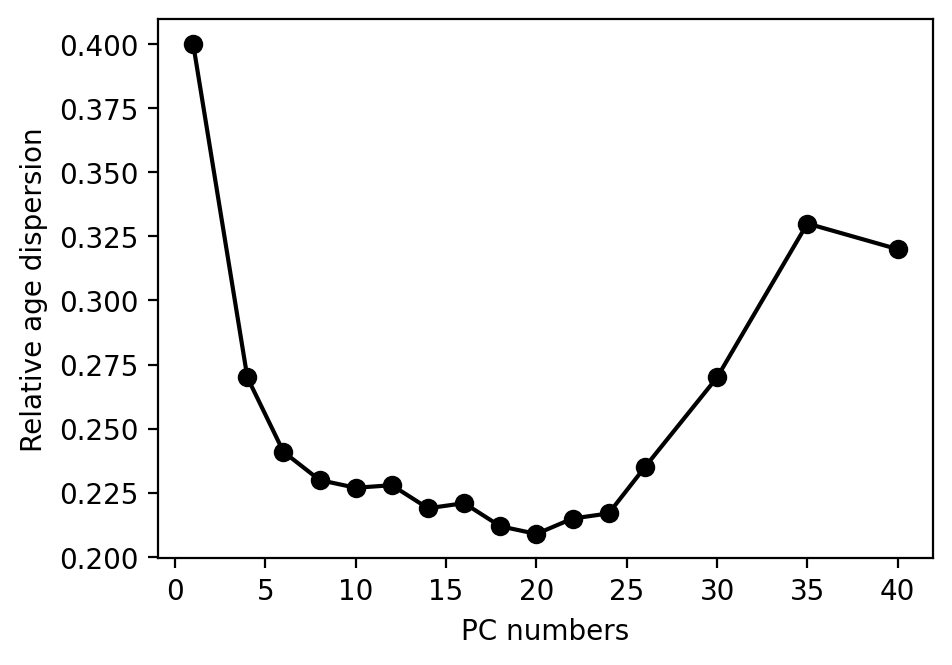}
\caption{The relationship between the number of PCs and the precision of age estimates.
\label{fig:npca}}
\end{figure}

\begin{figure*}[ht!]
\centering
\includegraphics[width=1\linewidth,]{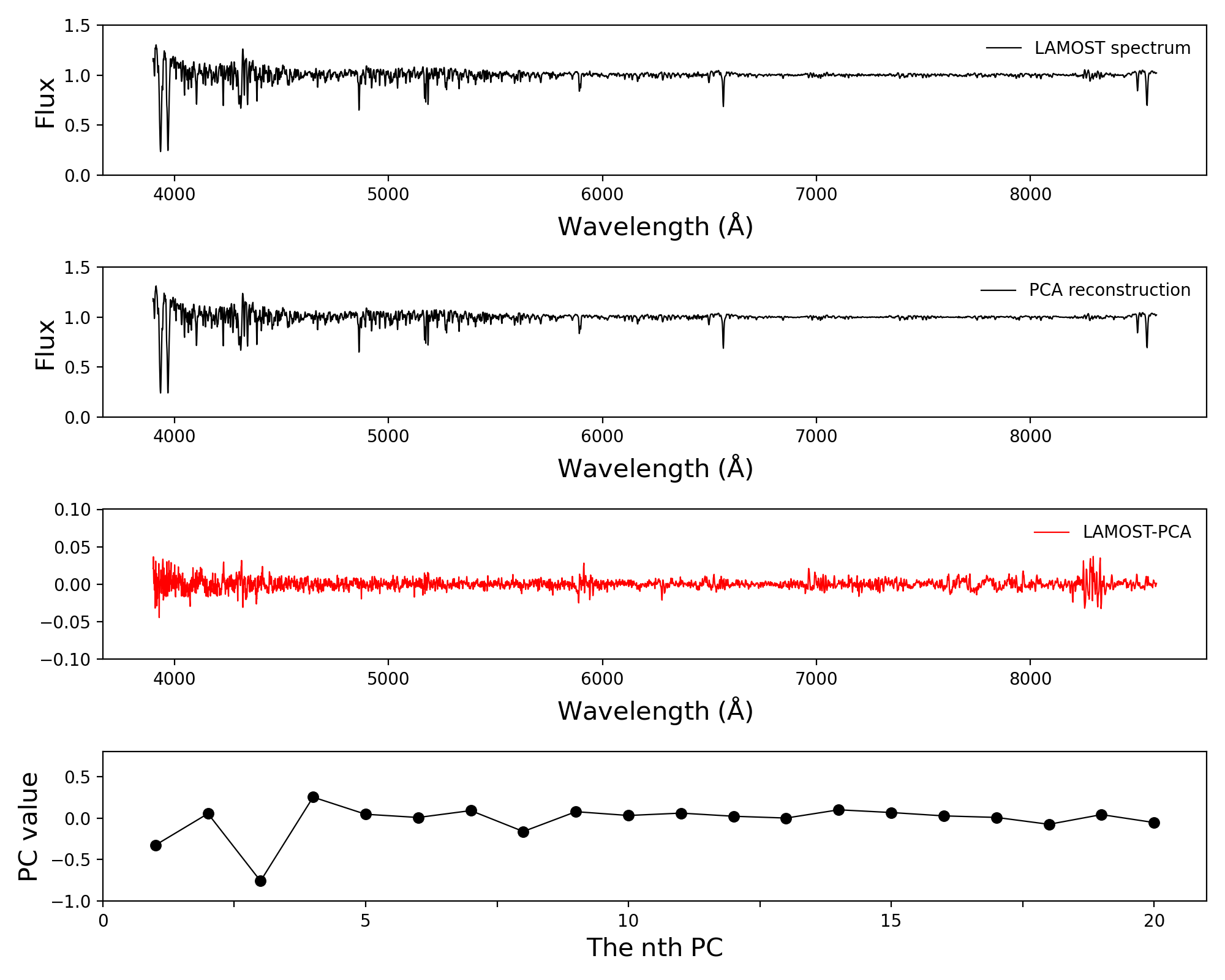}
\caption{An example for PCA reconstruction of the LAMOST spectra for de-noising. From top to bottom, the first panel is the normalized LAMOST spectrum. The second panel is the PCA-reconstructed spectrum. The third panel is the residual by subtracting the PCA-reconstructed spectrum from the LAMOST spectrum. The fourth panel is the value for the first 20 principal components (PCs). 
\label{fig:pca}}
\end{figure*}

We use the full sample of 2857 spectra to compute the principal components (PCs) and choose the first 20 PCs for the spectra reconstruction. Here the number of PCs adopted is determined by a brute-force search approach.  Specifically, we examine a series of PC numbers from 1 to 40, and implement the age regression on a validation set and calculated the relative age dispersion to determine the optimal number of PCs. As shown in Figure \ref{fig:npca}, using 20 PCs yields the best performance, and the precision of the age estimates no longer improves with increasing PC numbers. We then use the PCA spectra for age regression. Figure \ref{fig:pca} shows an example of the PCA reconstructed spectra as well as the noise spectra (residuals). 

\subsection{Validating the age estimation with validation set}
We divide our age sample into training and validation sets. Eighty percent of the wide binary stars and field stars are adopted as the training set, while the remaining 20\% are taken as the validation set. For each of the OCs, half of the member stars are used for training and the other half for validation. As the number of training samples is limited, the {\sc XGBoost} model training may suffer non-negligible uncertainty due to too sparse sampling. We therefore perform 20 training realizations with a bootstrap approach. In each realization, we randomly select the training and validation sets. Age predictions from the 20 training realizations are combined by taking the average value to minimize possible method error due to the training process. Specifically, the average age for the $i$th star in training (validation) set is defined as follows:

\begin{equation}\label{eq44}
\bar{\tau}_{\text{pred},i} = \frac{1}{N_i} \sum_{j=1}^{N_i} \tau_{\text{pred},i,j}
\end{equation}

$N_i$ is the number of realizations that the $i$th star is used for training (validation), which ranges from 2 to 20 as we require at least two realizations to compute average age. $\tau_{\text{pred},i,j}$ denotes the predicted age of the $i$th star using the $j$th training realization. It is worth noting that for the final LAMOST DR10 age catalog (see Section 4.1), the ages are averaged over 20 realizations for all stars. 

\begin{figure*}[ht!]
\centering
\includegraphics[width=1\linewidth,]{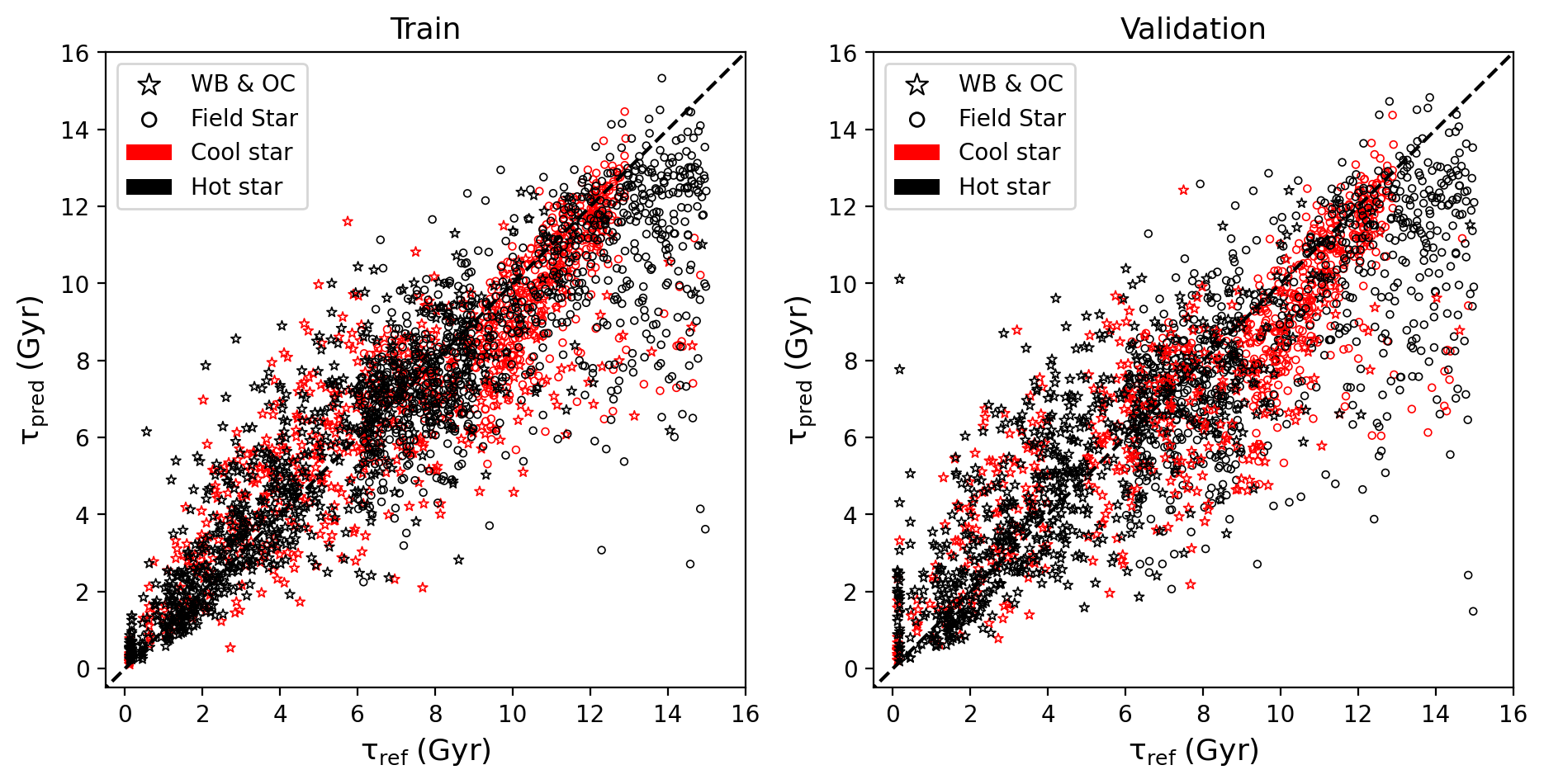}
\caption{Comparison of the XGBoost age estimates and reference (ground-truth) ages for our samples. The left panel shows results for the training sample, while the right panel shows results for the validation sample. Note the vertical axis refer to the averaged ages from different traing realizations (see text). Different tracers, namely wide binaries and open clusters (WB \& OC), and field stars, are shown different samples as marked in the Figure. Different colors represent different temperatures, with red for $T_{\rm eff} < 5400$K and black for $T_{\rm eff} > 5400$K. The dispersion in the relative age difference is found to be 15\% for the training sample and 18\% for the validation sample.
\label{fig:train}}
\end{figure*}

Figure \ref{fig:train} shows the comparison of the $\bar{\tau}_{\text{pred},i}$ with the ground-truth values for both the training and validation sets. The figure shows a good agreement between the predicted age and the ground-truth age, except for the youngest and oldest end. For the validation set, the youngest stars ($<0.5$~Gyr) may be overestimated significantly, while quite a number of the oldest stars ($\gtrsim13$~Gyr) are significantly underestimated (a similar effect is also found for the training set). However, the agreement on the whole is remarkable: the relative age dispersion for the overall sample is 15\% for the training sample, and 18\% for the validation sample. Given these two numbers are close to each other, we believe there is no significant overfitting in our model training process. 

As our training set is a mix of several sources, we found that in the validation set the relative age dispersion is $\sim$23.7\% for wide binary, and $\sim$12.5\% for field stars. The latter is smaller because we have set much more strict quality cut, in terms of both spectral $S/N$ and age precision, for selecting the training and validation set, as described in Sect. 2.2. When dividing the validation set in hot ($T_{\rm eff} > 5400$K) and cool samples, we found that the relative age dispersion is $\sim$19\% for the hot sample, and $\sim$15\% for the cool sample. The latter is lower because of two main reasons: first, the hot sample contains the majority of open cluster member stars, whose ages are young and thus have large relative errors (Sect. 5.1). Second, the cool sample contains the vast majority of old, cool high-alpha stars, which have good age estimates through the tight abundance-age relationship. After removing these old, cool high-alpha stars, the age dispersion for the cool sample becomes 23\%.

\subsection{Age error estimation}
To assign an error to each age estimate, we perform an Monte Carlo (MC) analysis by adding random Gaussian errors to the original spectrum and deriving the age repeatedly. Thirty iterations are implemented for each spectrum, and the standard deviation of the individual age estimates is adopted as the error estimate. Such an analysis takes the measurement errors of the spectral flux into account, thus giving a statistical uncertainty of the age estimate. A complete treatment of the age uncertainty should also consider other error sources, such as intrinsic error of the method, systematic age errors in the training set, etc. We will examine possible systematic errors in the age estimates in Section \ref{sec:result}.

\section{Result}\label{sec:result}
We applied our data-driven model to all low-resolution spectra of dwarf and subgiant stars in LAMOST DR10.In this section, we present statics of this large stellar age sample, and explore the correlation between age and other stellar physical quantities, including elemental abundances and activity.

\subsection{The LAMOST DR10 stellar age sample}
\begin{figure}[ht!]
\centering
\includegraphics[width=1\linewidth,]{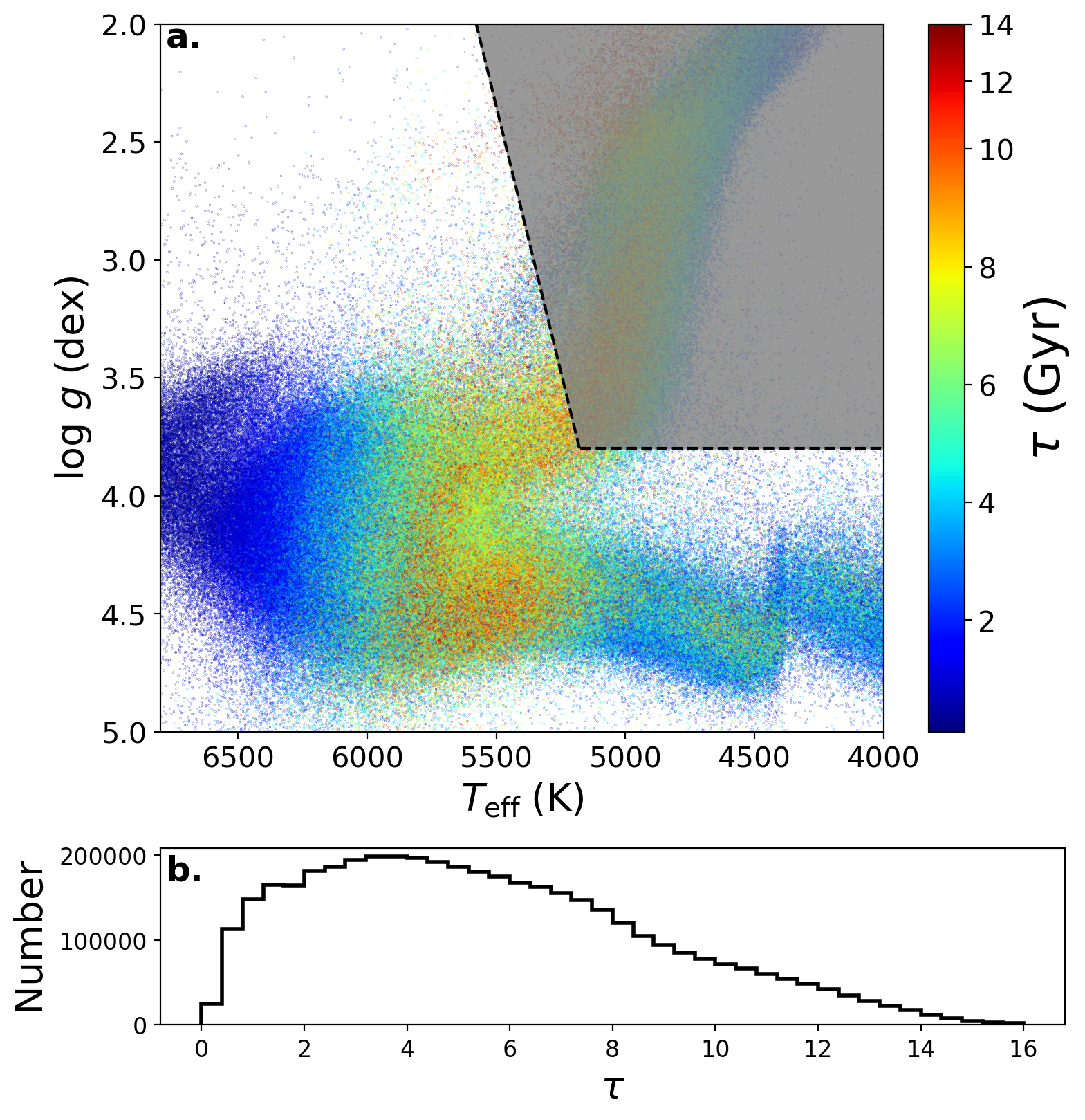}
\caption{Distribution of the LAMOST star sample. The upper panel (a) shows the distribution in the $T_{\rm eff}$-$\log~g$ diagram, color-coded by the stellar age estimates. To ensure precise age estimation, here only stars with $S/N>50$ are presented. The shaded area masks red giant stars that are not the focus of the current work. The bottom panel (b) shows the age distribution of the full stellar age catalog, with a total number of 4,233,948 stars.
\label{fig:allage1}}
\end{figure}

Owing to limitation of the training set, here we confine the analysis only to dwarf and subgiant stars. Therefore, we exclude red giants from the age sample, and the selection of red giants adopts the following criteria:
\begin{align}
&\log~g \lesssim 3.8 \\
&\log~g \lesssim -0.0045 \times T_{\rm eff}+27.1
\end{align}
as shown in the shaded window in Figure~\ref{fig:allage1}. The $T_{\rm eff}$ and $\log~g$ are from the updated LAMOST DR10 catalog (Mou et al. in prep.). In addition, we restrict the $T_{\rm eff}$ to the range of 4000–6800 K, in accordance with the coverage of the training sample. These criteria lead to a catalog of 3,868,513 stars in our resultant age catalog \footnote{The catalog are publicly available via https://zenodo.org/records/15824429}

\begin{figure*}[ht!]
\centering
\includegraphics[width=1\linewidth]{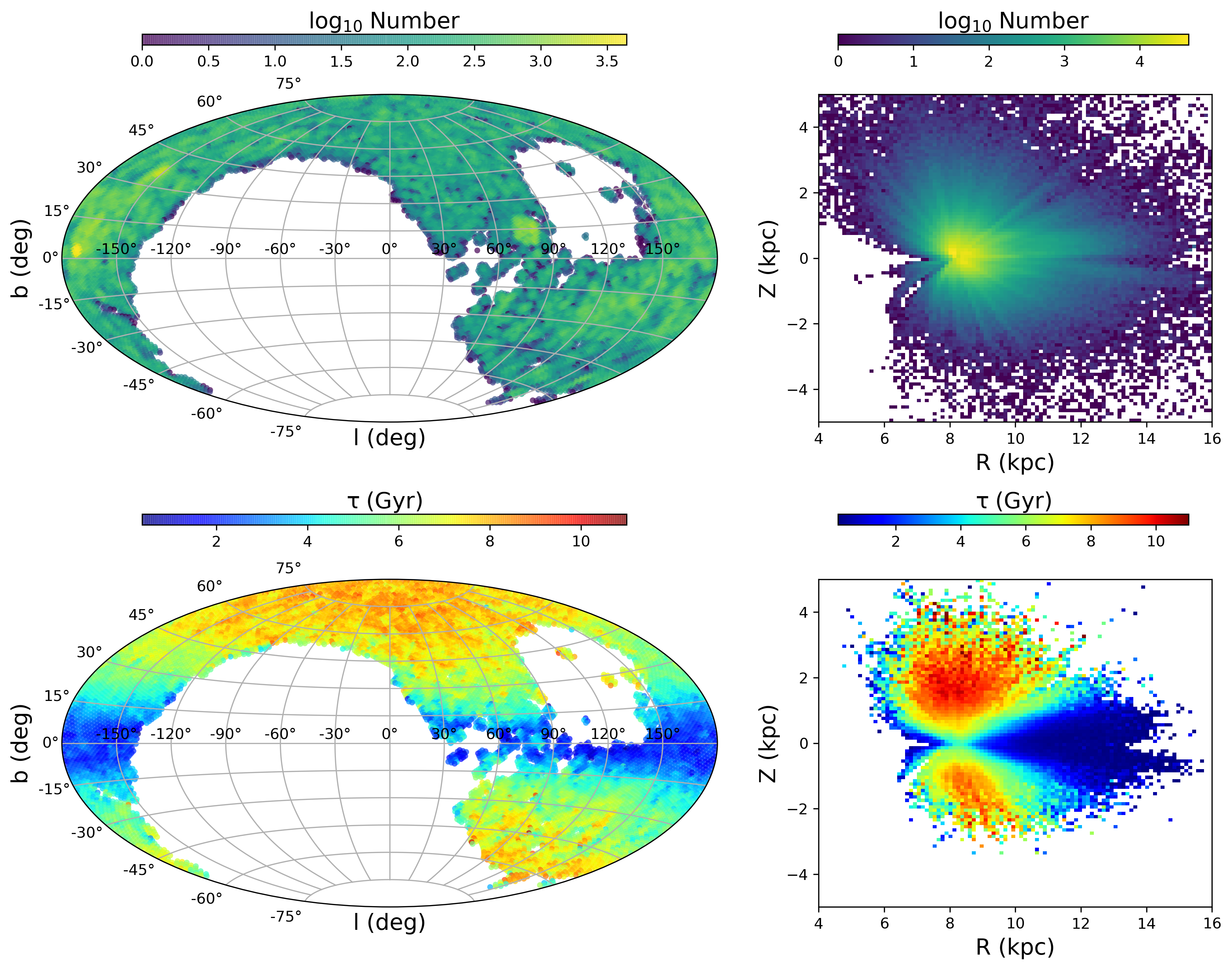}
% \plotone{binary.png}
\caption{Spatial density and age distribution of our dwarf star sample. The left panels show the stellar distribution in the Galactic coordinate system ($l$,$b$), color-coded by the stellar density (upper-left panel) and median age (lower-left panel). The right panels show the stellar distribution in the $R$-$Z$ plane of the Galactic cylindrical coordinate system, color-coded by the stellar density (upper-right panel) and median age (lower-right panel). Only stars with a spectral S/N higher than 30 are used to generate this figure.
\label{fig:age_space}}
\end{figure*}

Figure~\ref{fig:age_space} shows the spatial distribution of the number density and the age trend of the remaining sample. The upper panels of the Figure shows that our sample covers a large portion of the Galactic halo and anti-center disk, with a particularly dense sampling of the nearby disk ($8\lesssim R\lesssim10$~kpc). The mean stellar age shows a clear increasing trend with Galactic latitudes (bottom-left panel), and the age distribution in the $R$-$Z$ plane (bottom-right panel) of the disk exhibits a negative radial age trend, along with a strong flaring phenomenon. These features are consistent well with previous work \citep{2016ApJ...823..114N, 2016MNRAS.456.3655M, Xiang_2017, 2018MNRAS.481.4093S, 2019MNRAS.484.5315W, 2020ApJS..249...29H, 2022A&A...658A..91A}. 

The bottom-right panel of Figure~\ref{fig:age_space} also shows an unexpected trend in the inner disk of $R<8$ kpc that younger ages appear to occur at higher places (larger $Z$) as $R$ decreases. This is in contrast with the expectation that the inner and higher disk is older, reflecting an earlier, turbulent assembly history of the Galactic formation. However, as has been discussed in \citet{Xiang_2017}, this is an artifact due to the selection effect. Young stellar populations are intrinsically more luminous so that they probe to a larger distance along each sightline of the survey. Such an effect is especially severe for the inner disk because in those sky areas usually only bright stars (VB plates; $r<14$~mag) are targeted by the LAMOST survey. An accurate mapping of the age distribution requires a detailed correction of the survey selection effect, which is beyond the scope of this paper.

\begin{figure}[ht!]
\centering
\includegraphics[width=1\linewidth]{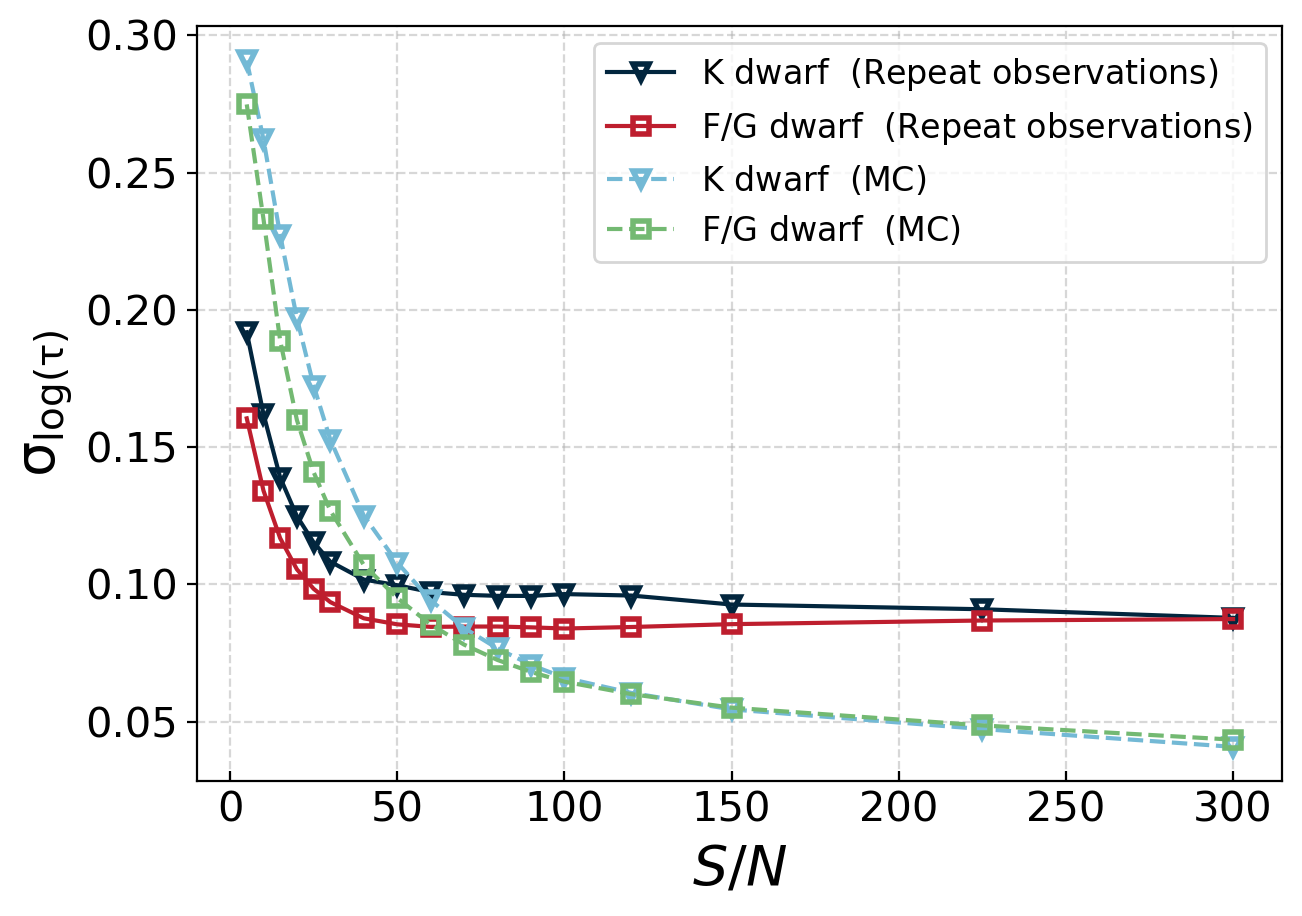}
% \plotone{binary.png}
\caption{Measurement error of the logarithmic age as a function of $S/N$. Results for F/G (5300-7000~K) and K dwarf (4500-5300~K) stars are shown in different symbols, are labeled in the Figure. Both measurement error estimated with a Monte-Carlo approach (see text) and dispersion among repeat observations are presented, as labeled in the Figure. 
\label{fig:repeatsn}}
\end{figure}

A large fraction ($\sim30\%$) of spectra in our sample are repeat observations of common stars. As a validation of the age estimates, Figure~\ref{fig:repeatsn} shows the dispersion of $\log$(age) between repeat observations as a function of spectral $S/N$. Here for repeat observation samples, we restrict to those that have comparable $S/N$ ($0.8<\frac{S/N_1}{S/N_2}<1.2$). The dispersion is then computed with a Gaussian fit to the $\log$(age) difference between the repeat observations, and divided by $\sqrt{2}$. The figure shows a dispersion decreasing from $\sim$0.18 (41\% in linear space) at $S/N=5$ to $\sim$0.08 (18\% in linear space) at $S/N>100$, slightly depending on the spectral type. F/G-type stars with higher temperatures have lower dispersion than the cooler, K dwarfs, which is expected as the ages of K dwarfs are harder to estimate due to their more crowded distribution in the $H-R$ diagram, which means that their locations in $H-R$ diagram have less sensitivity to their ages.

\subsection{Ages of star clusters}
\begin{figure*}[ht!]
\centering
\includegraphics[width=1\linewidth]{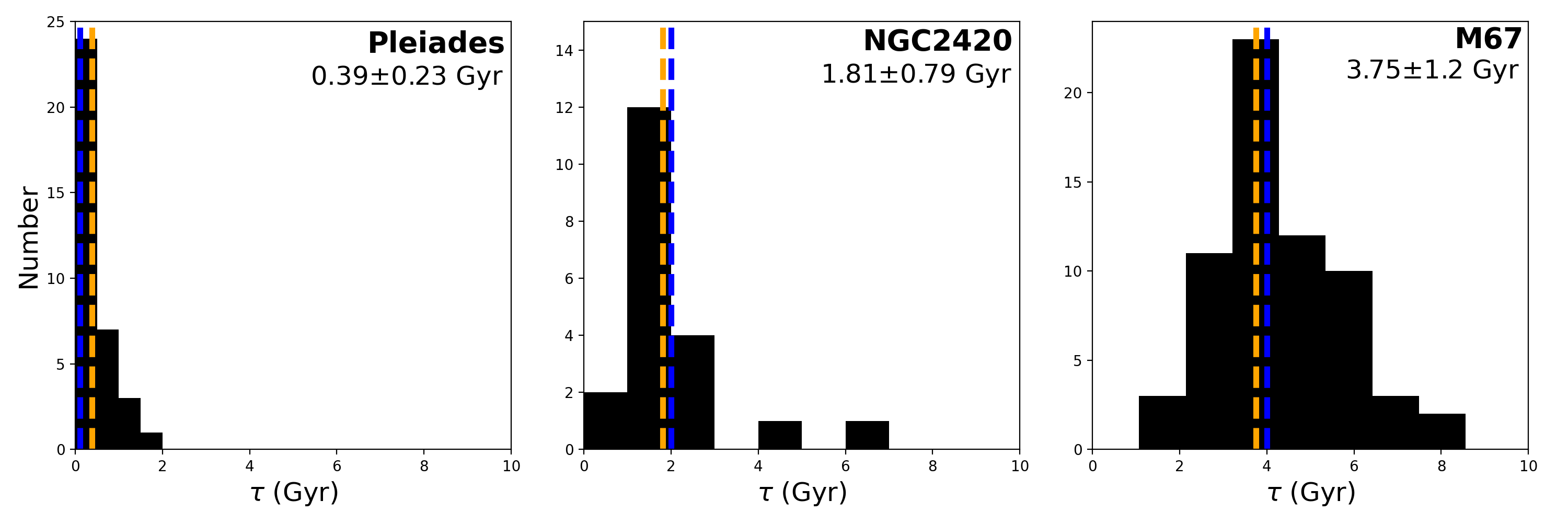}
\caption{Age distribution of cluster member stars. From left to the right panels are the results for Pleiades (M45), NGC2420, and M67 (NGC2682). Median age and standard deviation of the individual member stars are marked in the Figure. The vertical orange line represents the median age, and the blue line represents the literature age.
\label{fig:allage}}
\end{figure*}

The ages of star clusters provide an independent validation for our method. Here we present the results for three open clusters targeted by the LAMOST survey, namely, Pleiades (M45), and NGC~2420, M67 (NGC~2682). We use Gaia astrometric data to select cluster member stars based on their spatial positions (RA, Dec) and proper motions (PMRA, PMDec). In doing so, we first select LAMOST stars within a given radius around the cluster's center, and characterize the distribution of proper motion for the remaining stars with Gaussian fit. Stars deviating from the mean values of the Gaussian fit by $3\sigma$ are discarded as outliers. We restrict the analysis to stars that have a spectral $S/N$ higher than 30. The resultant age distribution is shown in Figure~\ref{fig:allage}.

For Pleiades, we adopt a cluster center of (RA, Dec)=($56.59359^{\circ}, +24.12223^{\circ}$) and a radius of $220.2$ arc minutes. The mean proper motion is (PMRA, PMDec) = (19.88, -45.41)mas/yr, and the dispersion ($\sigma_{\rm PMRA}$, $\sigma_{\rm PMDec}$) = (1.24, 1.43) mas/yr. This leads to 35 member stars in our sample, all of which were only used in the validation set while training the {\sc XGBoost} model. The left panel of Figure~\ref{fig:allage} shows that their age distribution exhibits a peak at 0.39~Gyr, with a standard deviation of 0.23~Gyr. This result is systematically older than literature values \citep[0.075--0.15 Gyr;][]{1996ApJ...458..600B} by 0.3~Gyr. This is likely due to an overestimate of our results for stars at the youngest border of the training set. For stars with ages close to or beyond the boundaries of the training sample, the XGBoost method tends to generate a lower age at the old end and a higher age at the young end. This is assumed to be a common bias effect for many machine learning methods \citep{2024arXiv241205806T}.

For NGC2420, we adopt a cluster center of (RA, Dec)=($114.6020^{\circ}, +21.5750^{\circ}$) and a radius of $19.8$ arc minutes. The mean proper motion is (PMRA, PMDec) = (-1.22, -2.05)mas/yr, and the dispersion $\sigma$ is (0.10, 0.09)mas/yr. This leads to 20 member stars in our sample. The middle panel of Figure~\ref{fig:allage} shows that the member stars spread an age between 1 and 3 Gyr, with a significant peak at 1.81~Gyr, and the standard deviation is 0.79~Gyr, corresponding to a relative age uncertainty of $\sim45\%$. The result is consistent with the literature, which is $2.0\pm0.2$~Gyr \citep{2000AJ....120.1384V}.

For M67, we adopt a cluster center of (RA, Dec)=($132.8460^{\circ}, +11.8140^{\circ}$) and a radius of $49.8$ arc minutes. The mean proper motion is (PMRA, PMDec) = ($-10.96$, $-2.91$)mas/yr, and the dispersion $\sigma$ is (0.22, 0.21)mas/yr. This leads to 80 member stars in our sample. The right panel of Figure~\ref{fig:allage} shows that the member stars have a median age of 3.75~Gyr, and a standard deviation is 1.2~Gyr, corresponding to a relative age uncertainty of $\sim30\%$. This result is also consistent well with the literature, which suggests 4~Gyr \citep[e.g][]{1998ApJ...504L..91R}.
 
\subsection{Age comparison with literature}
\begin{figure}[ht!]
\centering
\includegraphics[width=1\linewidth]{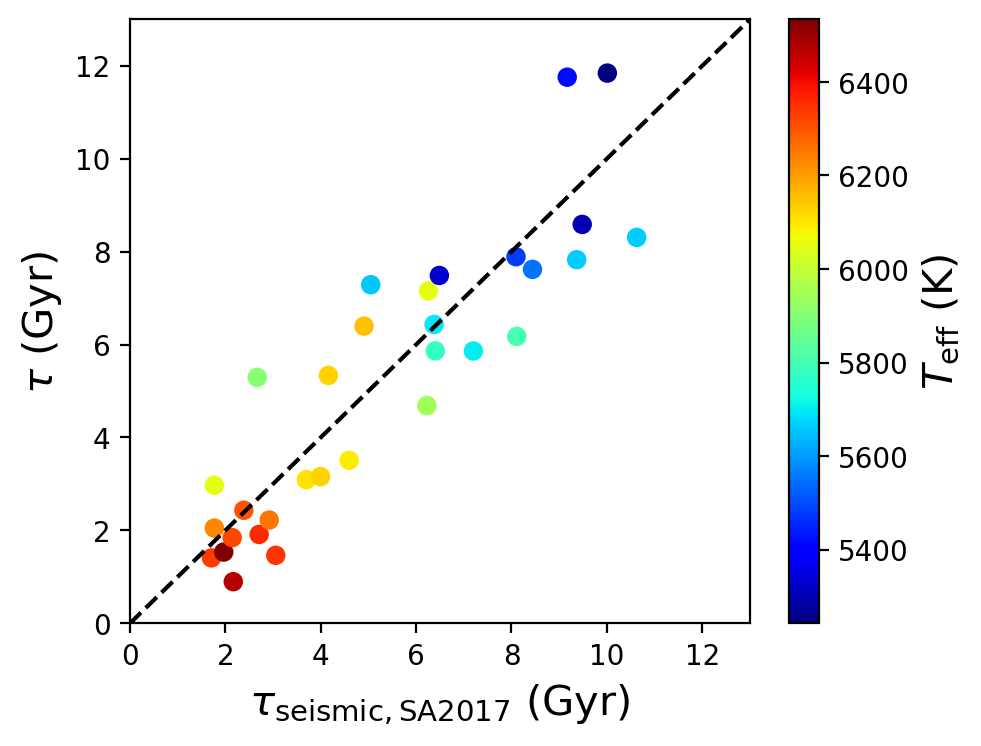}
% \plotone{binary.png}
\caption{A comparison of our age estimates with the asteroseismic ages of \citet{2017ApJ...835..173S}. Colors represent the effective temperatures from LAMOST spectra.
\label{fig:xie}}
\end{figure}

\citet{2017ApJ...835..173S} determined the ages  of 66 main-sequence (turn-off) stars utilizing asteroseismic and spectroscopic parameters as the constraints to stellar evolution models. Most of these stars have effective temperatures greater than 5400~K, and are close to the main-sequence turn-off in the HR diagram, while only 3 stars have effective temperatures below 5400~K. We cross-matched this asteroseismic age sample with our LAMOST DR10 age catalog, and obtained 31 common sources. \citet{2017ApJ...835..173S} employed seven different pipelines to calculate the ages, and the results among these pipelines showed a good consistency. Here we only compare our results with ages from their V\&A pipeline. As shown in Figure~\ref{fig:xie}, our age estimates show a good consistency with theirs, with a difference of approximately 20\%.

\begin{figure}[ht!]
\centering
\includegraphics[width=1\linewidth]{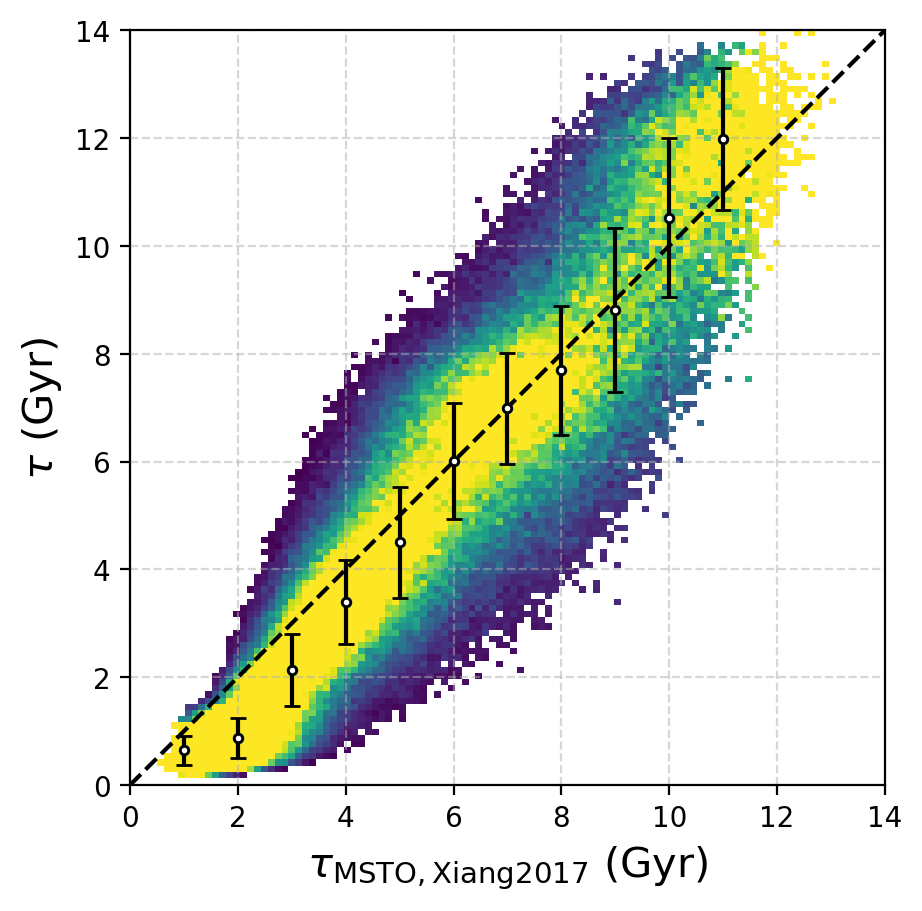}
% \plotone{binary.png}
\caption{A comparison of our age estimates with the isochrone ages of \citet{2017ApJS..232....2X}. Color represents stellar density normalized to the peak value row-by-row. The white points correspond to the median ages from 1 Gyr to 11 Gyr, with the error bars representing the dispersion.
\label{fig:xiangdr4}}
\end{figure}

\citet{2017ApJS..232....2X} provided the ages and masses of approximately 930,000 MSTO stars and subgiants in the Galactic disk from the LAMOST survey \footnote{http://dr4.lamost.org/v2/doc/vac}. They used a Bayesian algorithm to infer these stellar parameters by matching the effective temperature $T_{\text {eff }}$, absolute magnitude $\mathrm{M}_V$, metallicity [Fe/H], and $\alpha$ abundance [$\alpha$/Fe] with stellar isochrones. Here we compare our age estimates with \citet{2017ApJS..232....2X} for common stars with $S/N$ greater than 30. As shown in Figure~\ref{fig:xiangdr4}, the results exhibit a clear and tight consistency. The error varies with age, from approximately 0.3 Gyr at 1 Gyr to about 1.5 Gyr at 11 Gyr. We notice a slight systematic difference in the figure at the younger part ($\tau\lesssim4$~Gyr). Upon inspection, we suspect this is likely caused by the different stellar evolution models used, as age of the training sample in the current work is determined using the MIST isocrhone, while \citet{2017ApJS..232....2X} determined ages using the Yonsei-Yale (YY) stellar isochrones \citep{2001AAS...198.4301K,2004ApJS..155..667D,2024A&A...692A.243C,2024ApJ...976...87N}. These two isochrones adopt different schemes of core overshooting, which has a significant impact on the age determinations of young stars with intermediate mass.

\subsection{Age-activity relation}\label{sec:Aa}

\begin{figure*}[ht!]
\centering
\includegraphics[width=1\linewidth]{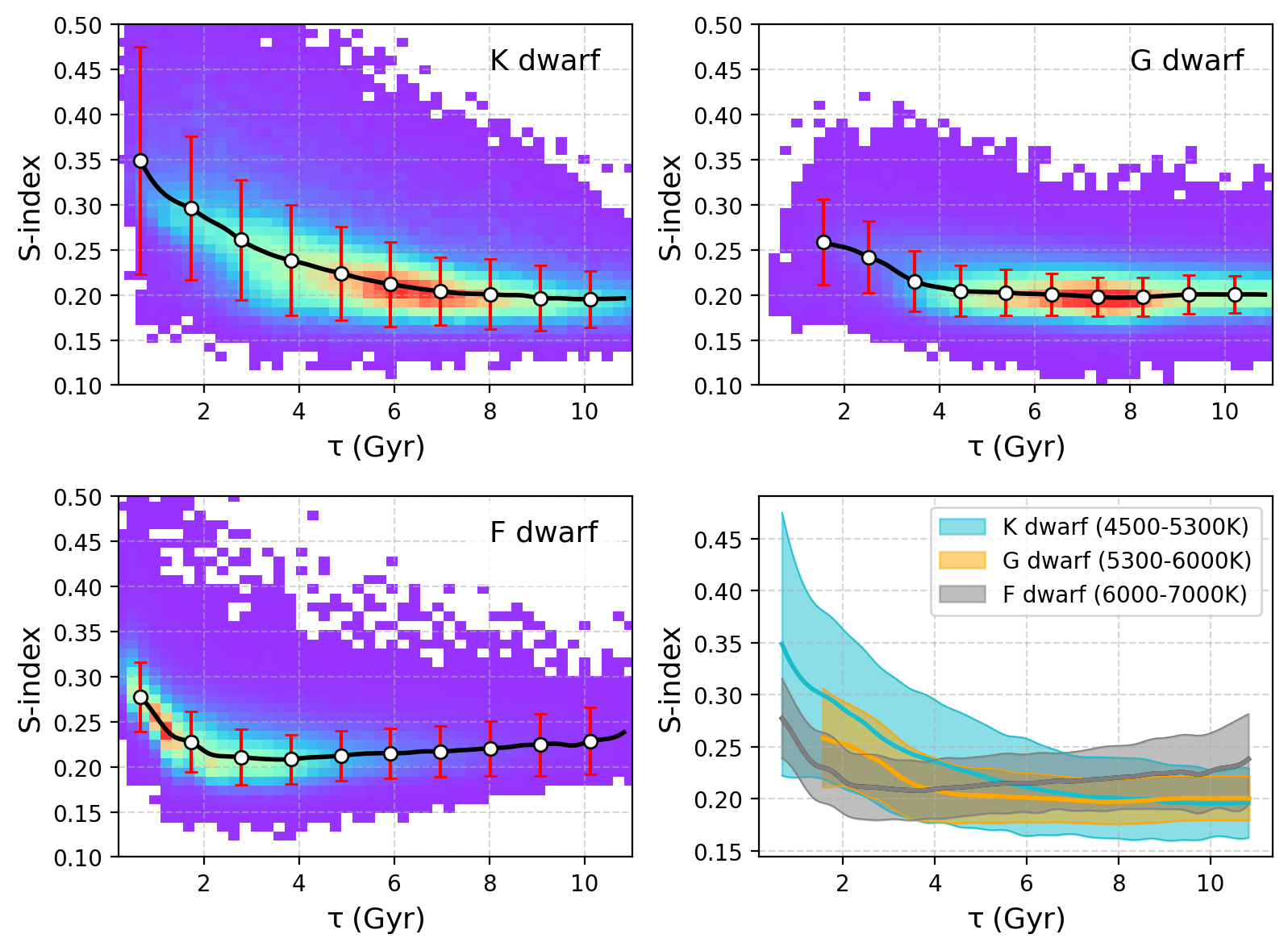}
% \plotone{binary.png}
\caption{Stellar activity S-index as a function of age for stars. The upper-left, upper-right, and bottom-left panels show the results for K dwarfs, G dwarfs, and F dwarfs, respectively. The dots and error bars are the mean and standard deviations of S-index in different age bins. The bottom right panel compares the results for these  F, G, and K dwarf stars.  
\label{fig:repeatsnr}}
\end{figure*}

It has been extensively discussed that the chromospheric activity of a star correlates with its age, and this relationship has been effectively utilized to estimate the ages of young dwarf stars. \citep[e.g.][]{1972ApJ...171..565S, 2010ApJ...722..222B,Soderblom_2010,2021AJ....162..100C,2024ApJS..271...19Y,2024ApJ...977..138H}. Here we investigate the relation between age and stellar activity with our sample. We adopt the stellar S-index from \citet[][]{2024ApJS..272...40Z} as the activity indicator. Figure \ref{fig:repeatsnr} presents the stellar S-index as a function of age for different types of stars in our sample. All of the F-(6000-7000K), G-(5300-6000K), and K-type(4500-5300K) dwarf stars exhibit a clear age-activity relation for relatively young stars, as younger stars show higher S-index. For old stars, such a relation however disappears for all of the F-, G-, and K-type dwarfs, and the S-index exhibits a flat distribution with age. These trends are qualitatively consistent well with literature \citep{Pace_2004,Zhao_2011}, validating the effectiveness of our age estimates.

Figure~\ref{fig:repeatsnr} also shows that the terminating age when the age-activity relation flattens varies with stellar types, from about 6~Gyr for K dwarfs to about 2~Gyr for F dwarfs. This indicates that as the temperature increases, the terminating age of dwarf stars moves toward the younger end. This feature is consistent with the prediction trend of theoretical evolution models.\citep{2015Natur.517..589M,2016Natur.529..181V}

\subsection{Relation between age and chemical abundances}\label{relationac}
In this section, we investigate the relation between the age and  $[\mathrm{X} / \mathrm{Fe}]$ as well as $[\mathrm{X} / \mathrm{Mg}]$. For demonstrating our age estimation method, here we focus on the K dwarf sample with $4800<T_{\rm eff}<5400$~K, whose ages are hard to be determined precisely with isochrone fitting (Section \ref{subsec:theoprecision}). We note that, as one expects our data-driven age estimates are largely constrained by features of stellar abundances in the spectra, i.e., the age estimates are correlated with the abundances, the age-abundance trends shown below are thus not independent measurements. Nevertheless, it is still valuable to look into these trends as an internal examination of possible intrinsic age-abundance trends, and is particularly informative on which chemical species contribute the majority of the age information that our method has implicitly utilized. 

The stellar abundances we utilized are derived from the LAMOST spectra with the {\sc DD-Payne}. While the early version of LAMOST {\sc DD-Payne} abundance catalog \citep{2019ApJS..245...34X} only contains elements with atomic numbers smaller than Zn (except for Ba), an updated version of {\sc DD-Payne} abundance catalog on LAMOST DR9 has included the abundances for a number of heavier elements, including Sr, Zr, Y, Ce, Ba, Nd, La, Sm, and Eu (Zhang et al. in prep.). We thus use this updated LAMOST DR9 abundance catalog for our analysis.    

\begin{figure*}[ht!]
\centering
\includegraphics[width=1\linewidth]{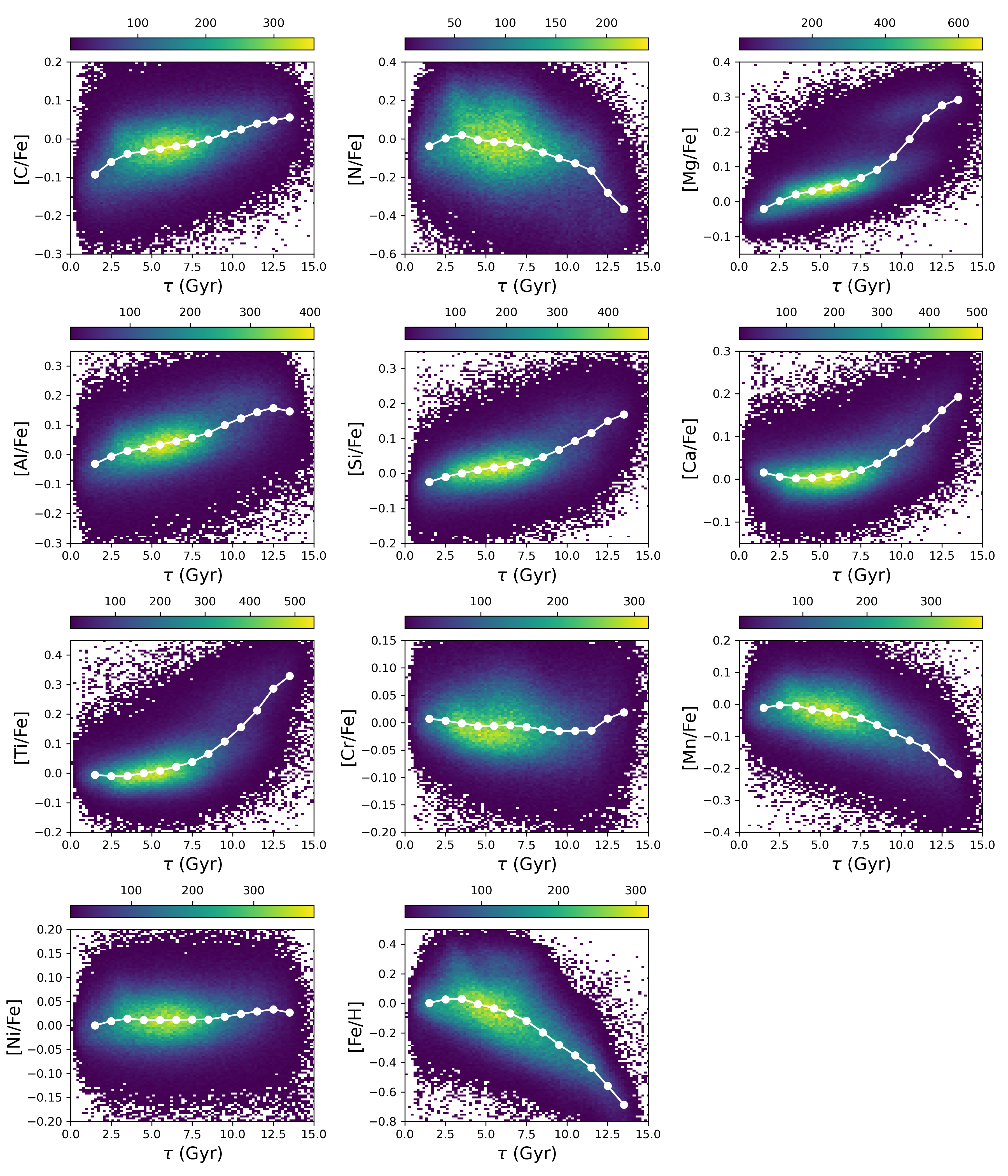}
% \plotone{binary.png}
\caption{Relations between age and elemental abundances for K dwarf stars. The abundances are derived from the LAMOST spectra with the DD-Payne (see text). The dots in white color are the median abundances in individual age bins.}  
\label{fig:age_abun}
\end{figure*}
\begin{figure*}[ht!]
\centering
\includegraphics[width=1\linewidth]{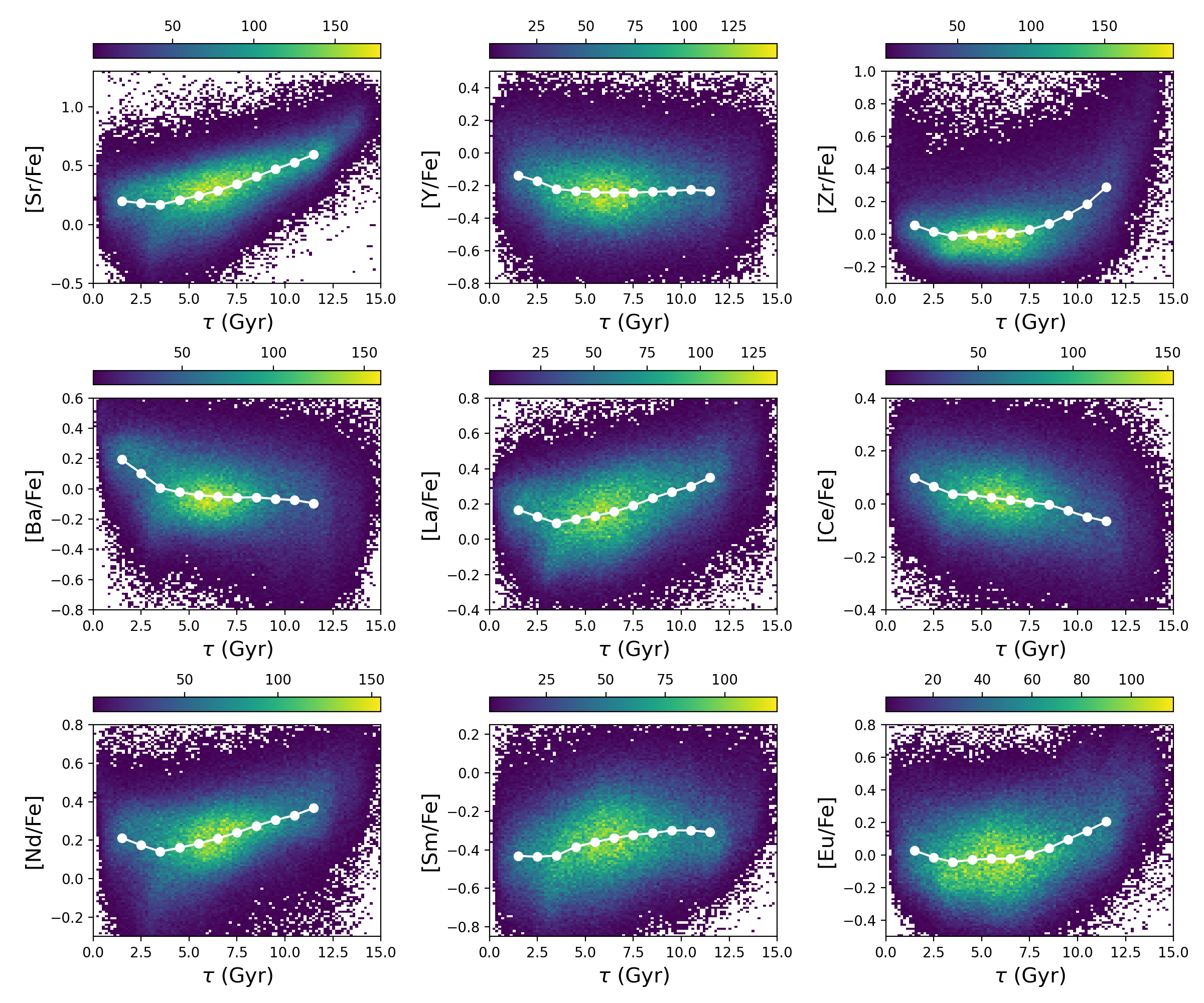}
\caption{Same as Figure~\ref{fig:age_abun}, but for abundances of heavier, s- and r-process elements.}
\label{fig:age_y_fe}
\end{figure*}

Figure~\ref{fig:age_abun} shows the stellar number density distribution in age-abundance planes. In particular, the figure shows a clear separated double-sequence feature in the age-[Mg/Fe] plot. Stars in the sequence with higher [Mg/Fe] have older ages, with ages older than 10~Gyr, while stars in the low-[Mg/Fe] sequence are systematically younger \footnote{see Section \ref{appendix1} in the Appendix for more details about the old stars in low-[Mg/Fe] sequence.}. Such a double-sequence trend is in good agreement with previous work \citep[e.g.][]{2018MNRAS.475.3633W}. 
Beyond this double-sequence feature, Figure~\ref{fig:age_abun} shows clear age-abundance correlations for most of the elements.
For stars older than 4~Gyr, the mean values of [C/Fe], [Mg/Fe], [Al/Fe], [Si/Fe], [Ca/Fe], and [Ti/Fe] show a monotonous decrease with decreasing age, the [Cr/Fe] and [Ni/Fe] show a nearly flat trend with age, while the mean values of [N/Fe], [Mn/Fe], and [Fe/H] show an monotonous increase with decreasing age. However, for stars younger than 4~Gyr, these trends are either flattened or reversed for some of the elements, such as [N/Fe], [Ca/Fe], [Ti/Fe], [Mn/Fe], and [Fe/H].   
%the age-[Fe/H] relation breaks up at 6~Gyr. Stars older than 6~Gyr exhibit a tight negative relation as older stars are more metal-poor, while the age-[Fe/H] relation becomes flat or even slightly positive for stars younger than 6~Gyr.
%\sout([N/Fe]), [Ca/Fe], [Ti/Fe], [Mn/Fe], albeit the age relations are positive for [Ca/Fe] and [Ti/Fe] at the old part. The [C/Fe], [Mg/Fe], and [Al/Fe], however, show a monotonous positive trend with age. The age correlations for [Cr/Fe] and [Ni/Fe] are weak for \sout{old stars but moderate for young}all stars. 

Similar age-abundance trends are also clearly presented for heavier, s- and r-process elements (Figure~\ref{fig:age_y_fe}). There is a strong positive age trend for [Sr/Fe], [La/Fe], and [Nd/Fe] for stars older than 4~Gyr, but the trends are reversed for stars younger than 4~Gyr. Albeit less strong, similar trends for [Zr/Fe], [Sm/Fe], and [Eu/Fe] are also visible. The [Y/Fe] and [Ba/Fe] show almost a flat or slightly negative trend for stars older than 4~Gyr, but exhibit a clear negative trend for satars younger than 4~Gyr. The [Ce/Fe], however, exhibit a negative age trend for all ages. 

% \begin{figure*}[ht!]
% \centering
% \includegraphics[width=1\linewidth]{age_y_mg.png}
% % \plotone{binary.png}
% \caption{Same as Figure~\ref{fig:age_y_fe}, but for [X/Mg]
% \label{fig:age_y_mg}}
% \end{figure*}

The age-abundance trends are qualitatively consistent with literature. Using data from the Gaia-ESO survey, \citet{2016A&A...589A.115S} found that [Al/Fe] was positively correlated with age, which is consistent with our results. \citet{2023MNRAS.523.1199S} obtained the same trend of [Al/Fe] using MSTO from the GALAH survey. The slightly negative trend with age for [Ba/Fe] is consistent with the results of high-resolution spectroscopy \citep{2019A&A...624A..78D}, and with the GALAH survey \citep{2022MNRAS.517.5325H,2022MNRAS.510..734S}. \citet{2022MNRAS.510..734S} also found that the age-[Eu/Fe] relationship showed a positive correlation trend, which is consistent with our result for stars old than $\tau >$ 4~Gyr. 
\citet{2024MNRAS.528.3464R} suggested that the age-abundance trend changes with the birth radius of the star. This is probably the reason for the broken age-abundance relations seen in our sample, as we expect the old stars in our sample to be mostly born at the inner disk, while the young stars ($\tau\lesssim4$~Gyr) are mostly born at the outer disk \citep[e.g.][]{2021MNRAS.501.4917W, 2022arXiv221204515L}. 

These well-presented age-abundance relations imply that the LAMOST spectra indeed contain a wealth of features that can be used to estimate the stellar age, at least from a chemical clock. The results also suggest that the {\sc DD-Payne} abundance estimates are robust, and can provide constraints on their ages.

\section{discussion}  \label{sec:discussion}
While we have shown that the {\sc XGBoost} regression approach can pragmatically estimate the age of cool main-sequence dwarf stars from the LAMOST spectra to an internal precision of 20-40\%, it is important to understand the underlying clocks that our method has utilized implicitly for estimating the age. This is especially essential for better understanding limitations in the age estimates. For this purpose, we have made the following experiments:
\begin{itemize}
    \item First, we examine the precision of age estimation with isochrone fitting method utilizing the stellar atmospheric parameters, i.e., $T_{\rm eff}$, $\log~g$, and [Fe/H], for a typical cool main-sequence star. A comparison of the precision between the isochrone age and the data-driven age provides insights into which extent the latter relies on clocks of stellar evolution.
    \item Second, we build chemical clocks to infer stellar age with LAMOST  abundances. A comparison of precision between the chemical age and the data-driven age tells to which extent the latter relies on clocks of Galactic chemical evolution. 
    \item Third, we inspect how the data-driven age estimation relies on stellar activity by masking activity-sensitive lines, including Ca~{\sc ii}~H\&K and H$\alpha$ lines.   
\end{itemize}

\subsection{Precision of isochrone ages for K dwarfs}\label{subsec:theoprecision}
Stellar age can be estimated from their atmospheric parameters via fitting to stellar isochrones. For convenience, let's denote this isochrone age estimate as, 
\begin{equation}\label{eq6}
\tau_{\rm iso}=f \left(T_{\rm eff},\log~g, {\rm [Fe/H]} \right).
\end{equation}
Then one may have an approximate estimate of the uncertainty in the age estimate via,  
\begin{equation}\label{eq7}
\delta_\tau=\sqrt{
\sum_i\left( \frac{\partial{\tau}}{\partial{X_i}}\right)^2\delta_{X_i}^2
}
\end{equation}
where $X_i$ refers to $T_{\rm eff}$, $\log~g$, and [Fe/H], $\delta_{X_i}$ refers to uncertainties in these atmospheric parameters.

\begin{figure}[ht!]
\centering
\includegraphics[width=1\linewidth]{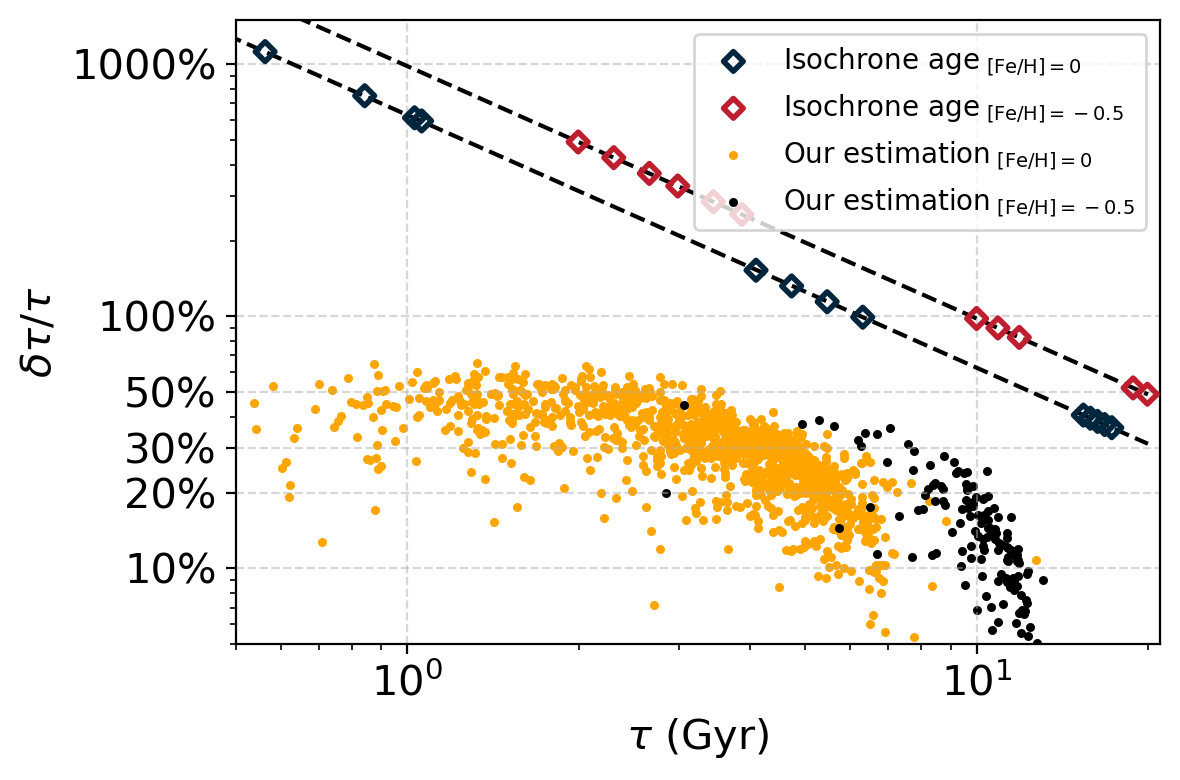}
% \plotone{binary.png}
\caption{Relative age error as a function of age for K dwarfs ($T_{\rm eff} = 5000$~K). The orange and black dots are age estimates for two different metallicities, as labeled in the Figure. As a comparison, the expected errors of isochrone age estimates for K dwarfs are shown in diamonds (see text). The dashed line is a linear fit to the diamonds.     
\label{fig:iso_err}}
\end{figure}

For $K$ dwarf stars with $T_{\rm eff}\simeq5000$~K, typical internal uncertainties in the LAMOST stellar parameters are $\delta T_{\rm eff}$ = 20K, $\delta \log~g$ = 0.05, $\delta \rm [Fe/H]$ = 0.07, given a spectral $S/N$ higher than 50. To have an estimate of $\delta_\tau$, we deduce a local derivative $\frac{\partial\tau}{\partial{X}}$ in Eq.\ref{eq7} for each fiducial star ${X_i}$ of concern using the MIST stellar isochrones. Figure~\ref{fig:iso_err} shows the resultant relative age error estimates for several fiducial $K$ dwarfs of different ages, selected from MIST isochrone grid points, at both ${\rm [Fe/H]}=0$ and ${\rm [Fe/H]}=-0.5$. The relative age error increases linearly with decreasing age in logarithmic space, from 50\% for stars of 10~Gyr to $\sim$1000\% for stars of 0.5~Gyr. This corresponds to a constant age error of 5~Gyr in absolute scale, owing to the nearly even distribution of stellar isochrones of different ages in the parameter space of concern, i.e., $\frac{\partial\tau}{\partial{X}}$ is nearly constant. 

Such an isochrone age error is apparently much larger than the data-driven age error we obtained in the current work, as the latter has a typical value of $\simeq$15\% for stars of 10~Gyr, and 30-50\% for young stars of $\tau\lesssim2$~Gyr (Figure~\ref{fig:iso_err}). On one hand, this clearly illustrates that isochrone ages for main-sequence $K$ dwarfs suffer huge uncertainties. On the other hand, it implies that stellar atmospheric parameters only have contributed a small portion of the information that our data-driven approach used for age inference, while most of the information for age inference must be from other stellar clocks, presumably chemical abundances (Sect.~\ref{subsec:5.2}). 

\subsection{Age estimates from chemical clocks} \label{subsec:5.2}
Section~\ref{relationac} has shown that there are tight correlations between our age estimates and the chemical abundances. These correlations may have served as chemical clocks that provide major information for the data-driven age estimation.

\begin{figure*}[ht!]
\centering
\includegraphics[width=1\linewidth]{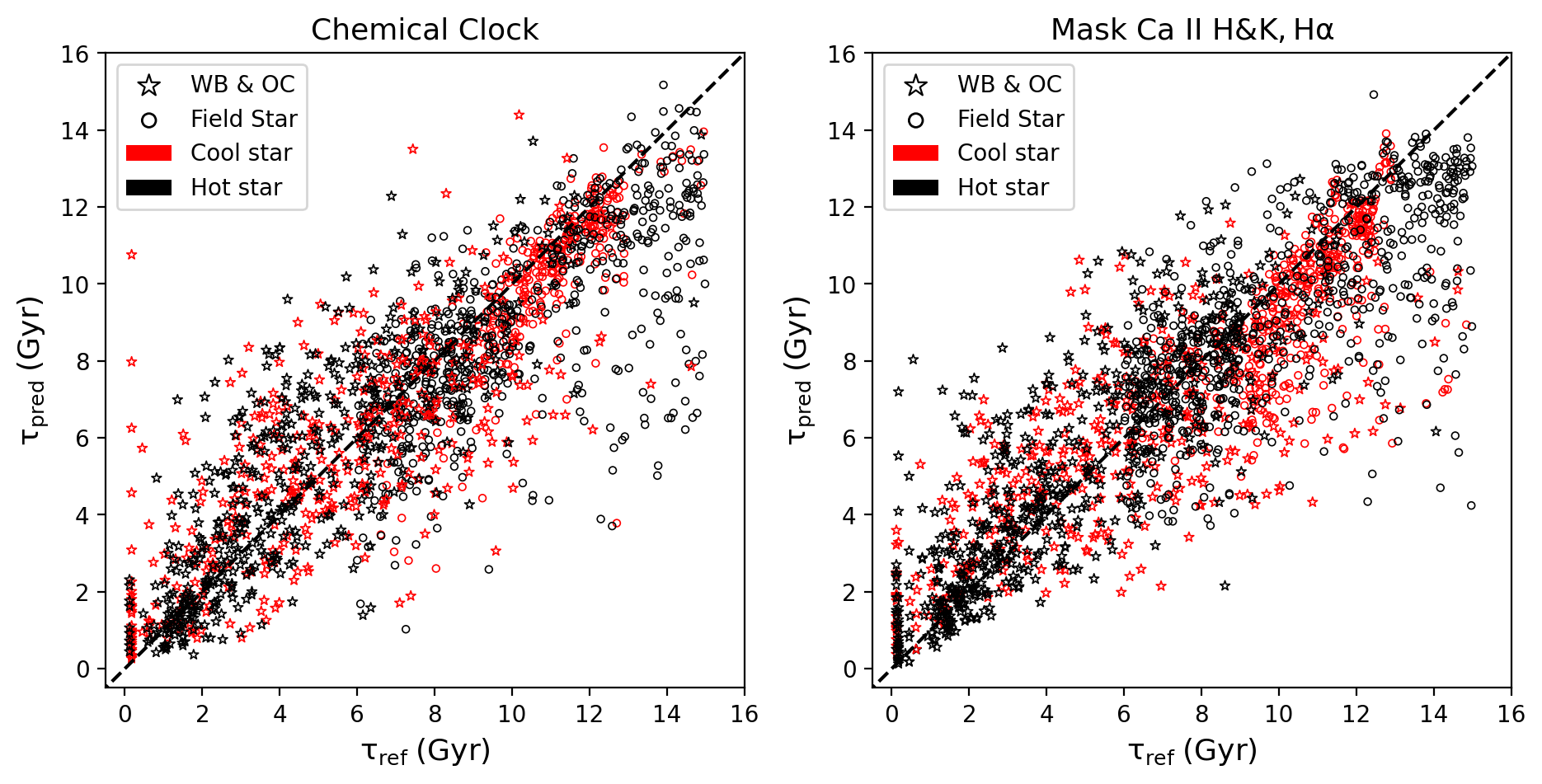}
% \plotone{binary.png}
\caption{Same as Fig. 6, but for stellar age estimates with alternative methods. The left panel shows the age estimated from an empirical chemical clock built in this work. The right panel shows the data-driven spectroscopic age estimates after masking Ca~{\sc ii} H, K lines, which may be affected by stellar activity (see discussions in Section~5).  }
\label{fig:train_abun}
\end{figure*}

To have a quantitative examination, here we derive age estimates directly from the chemical abundances by building an empirical chemical clock,
\begin{equation}
\tau_{\rm [X/H]}=f\left(T_{\rm eff},\log~g, {\rm [X/H]} \right),
\end{equation}
where [X/H] refers to 12 element abundances, namely, [Fe/H], [C/Fe], [N/Fe], [Mg/Fe], [Ca/Fe], [Si/Fe], [Ti/Fe], [Ba/Fe], [Cr/Fe], [Mn/Fe], [Ni/Fe] and [Al/Fe], derived with {\sc DD-Payne} from the LAMOST spectra. This empirical relation is again constructed from our training sample with the {\sc XGBoost} method. We then apply this chemical clock to our test sample. As shown in Figure~\ref{fig:train_abun}, the chemical age is in good consistency with the reference age, and the overall dispersion in the relative age difference is 17\%, almost identical to the data-driven age (18\%, Figure~\ref{fig:train}). Inspired by this comparison, in an upcoming work we will provide another set of stellar ages for the LAMOST stars, specifically derived with the chemical clock, using the abundances for more than 20 elements, including the heavier, s- and r-process ones.      
These results suggest that our data-driven approach of age estimation is largely identical to a chemical clock, that is, the information from which the data-driven approach infers the age estimates is mostly from spectral features of chemical abundances. This is particularly true for cool main-sequence dwarf stars, however, for hotter main-sequence or subgiant stars, the isochrone age may have comparable or even better precision to the chemical age, stellar evolution thus serves as the main clock for the current results. On the other hand, as discussed in Section \ref{relationac}, because our age estimates rely (implicitly) on chemical abundances, the age-abundance correlation from our sample is no longer an independent derivation. We emphasize that one needs to exercise great caution in any studies about the age-abundance relation using our age catalog.

\subsection{Contribution from stellar activity}
The LAMOST spectrum contains a few lines that are good indicators of stellar activity, including the Ca~{\sc ii} H and K lines, and the ${\rm H}_\alpha$ line. To examine how our data-driven age estimation depends on these features, we have made a test for the age estimation by using the LAMOST spectra windows after masking these activity-sensitive lines. Indeed, after performing PCA, much of the activity information in the spectra may have been removed. As shown in the right panel of Figure~\ref{fig:train_abun}, the resultant age estimates are almost identical to those from the full spectra, meaning that these activity-sensitive lines contributed only a negligible portion of the information for the data-driven age estimation. The stellar age-activity relations presented in Sect.~4.4 are therefore reliable. At the same time, this also implies that the chemical information in the spectrum is enough to constrain age measurements, and the effect of chemical tagging is no less than that of activity information, even for young stars.

It has been shown that there are tight correlations among age, stellar activity, and stellar intrinsic photometric noise \citep{2024ApJS..272...40Z, 2024ApJS..271...19Y}. Our age catalog thus provides a piece of key information to select quiet stars with low-level intrinsic photometric noise for planet detection with transit observations, such as the coming Earth 2.0 mission \citep{2024AAS...24344501G}. 

\subsection{Limitation of the age estimates}

While our method has achieved robust age estimates for given feasible spectral $S/N$, one has to be in cautious about some caveats in the age estimates. First of all, as mentioned above (Sect. 4.2), while we can distinguish between young and old stars, there maybe a systematic error in absolute age estimation at the youngest end ($\textless$1Gyr). We believe that the systematic error maybe insignificant only for stars older than about 1~Gyr. In addition, the age needs to be used with caution in the following specific cases: i) the age estimates can be unreliable for binary and multiple star systems; ii) the age estimates can be incorrect for stars accreted from dwarf galaxies, which have different chemical evolution histories compared to native stars. This is a consequence of the fact that our training sample is not constructed to contain a significant number of stars from dwarf galaxies; iii) our age estimates are assumed to be inapplicable for pre-main sequence stars, as these stars may have a different age-spectra relation compared to main-sequence stars.

\section{conclusion}  \label{sec:conclusion}

It is challenging to determine the age of cool main-sequence dwarf stars accurately. In this work, we have employed a data-driven method to estimate age of dwarf stars from the LAMOST low-resolution spectra. We take 1213 wide binaries of main-sequence dwarf plus subgiant systems, 927 warm field stars and 400 cool dwarfs in the old disk with known age estimates, and 317 cluster member stars as the training set. For wide binaries, the age of the cool main-sequence dwarf companion is known from its warmer primary, whose age is determined via isochrone fit. We adopt the {\sc XGBoost} algorithm to train the data-driven model for age estimation from the LAMOST spectra. Our main results are listed below: 
\begin{itemize}
    \item Given a LAMOST spectral S/N higher than 50, our method achieved a typical precision of 0.1~dex in the $\log{\rm Age}$ estimates for K-type dwarf stars, corresponding to $\sim$23\% relative age uncertainty in linear scale, while the uncertainty increases to 0.15~dex at $S/N\simeq20$. Additionally, the relative age uncertainty also varies with age. For K-dwarfs, it is 50\% at 1 Gyr and decreases to 15\% at 10 Gyr.
    \item We found that for cool main-sequence stars ($T_{\rm eff}\lesssim5400K$), chemical abundances in the spectra provided most of the information for the data-driven age inference, while other information such as stellar parameters ($T_{\rm eff}$, $\log~g$, [Fe/H]) and stellar activity indicators (Ca~{\sc ii} H and K lines, ${\rm H}_\alpha$ line) contribute little to the age inference. As a result, one needs to be very cautious in using our age estimates for these stars for studying the stellar age-abundance correlations.
    \item Our results reveal a clear age-stellar activity relation for relatively young stars ($\tau\lesssim5$~Gyr), with a detailed trend varying with spectral types, while older stars essentially show a flattened age-activity relation. 
    \item We applied the method to the full dataset of LAMOST DR10, and obtained the age estimates for about 4 million stars. This age catalog is expected to be helpful for i) the study of star formation and assembly history of our Galaxy; ii) the study of temporal variations of the stellar initial mass function (IMF); iii) the selection of quiet stars with low intrinsic photometric noise for planet detection through transiting observations, such as that on the coming Earth 2.0 mission.
    
    \item The current sample of wide binary stars in our training set is limited. We expect that the ongoing LAMOST Phase-III survey will significantly increase the spectroscopic wide binary sample, and our method can therefore achieve a better age estimation in the next data release. 

\end{itemize}

We acknowledge financial support from the NSFC through grant Nos. 2022000083 and the National Key R\&D Program of China through grant No. 2022YFF0504200. J.-H.Z. acknowledges support from NSFC grant No. 12103063.

This work has made use of data products from the Guo Shou Jing Telescope (the Large Sky Area Multi-Object Fiber Spectroscopic Telescope; LAMOST). LAMOST is a National Major Scientific Project built by the Chinese Academy of Sciences. Funding for the project has been provided by the National Development and Reform Commission. LAMOST is operated and managed by the National Astronomical Observatories, Chinese Academy of Sciences.
% \section{Acknowledgments
% \begin{nolinenumbers}
% \begin{acknowledgments}
% We acknowledge financial support from the NSFC through grant Nos. 2022000083 and the National Key R\&D Program of China through grant No. 2022YFF0504200. J.-H.Z. acknowledges support from NSFC grant No. 12103063.

% This work has made use of data products from the Guo Shou Jing Telescope (the Large Sky Area Multi-Object Fiber Spectroscopic Telescope; LAMOST). LAMOST is a National Major Scientific Project built by the Chinese Academy of Sciences. Funding for the project has been provided by the National Development and Reform Commission. LAMOST is operated and managed by the National Astronomical Observatories, Chinese Academy of Sciences.
% \end{acknowledgments}
% \end{nolinenumbers}

\appendix

\section{Justification for the age estimation with old Mg-poor stars} \label{appendix1}
We have noticed that in the $\tau$–[Mg/Fe] relation plot, there are some old Mg-poor stars $(\tau \textgreater \rm 10~Gyr)$. To validate their age estimation, we carry out a more detailed analysis on these stars. We choose low-alpha K dwarf stars located in a small box in the $\tau$-[Mg/Fe] and [Mg/Fe]-[Fe/H] planes, as highlighted in red dots in the left and middle panels of Figure \ref{fig:a1} below, and plot their age as a function of temperature in the right panel. As a comparison set, we also show the result of stars with similar [Fe/H] and [Mg/Fe] but with higher $T_{\rm eff}$ ($\textgreater 5400$ K). The ages of the comparison stars are plotted with white dots in the right panel of Figure \ref{fig:a1}. Their ages show a clear trend with temperature, spanning from 6 to 12 Gyr in the temperature range 5400-6000 K. Most stars with $T_{\rm eff}~\textless$ 5600 K have an age older than 10 Gyr. We expect that a substantial portion of the information for their age estimation is $T_{\rm eff}$, $\log~g$, and [Fe/H], and their age estimates thus are robust. Otherwise, we would expect these stars to have a more uniform age distribution, similar to the background rather than concentrate on 10–12 Gyr. The K dwarfs (red dots) are well on the $\tau-T_{\rm eff}$ trend, and we thus believe their age estimates are reasonable. In particular, we do not expect there to be an abrupt difference in the age distribution between $T_{\rm eff}~\textless$ 5400 K and $T_{\rm eff}~\textgreater$ 5400 K for stars in the same [Fe/H] and [Mg/Fe] bin.

\begin{figure*}[ht!]
\centering
\includegraphics[width=1\linewidth]{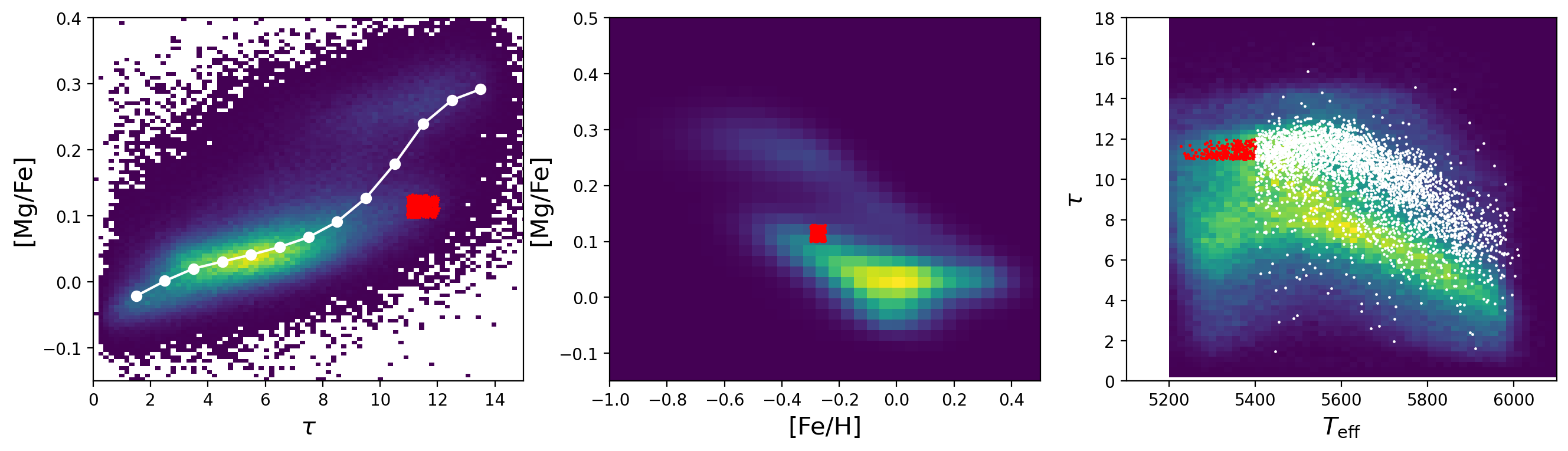}
% \plotone{binary.png}
\caption{Validation of the age estimates for old Mg-poor stars. The left panel shows the $\tau$–[Mg/Fe] relation for K dwarfs, from which we have selected old Mg-poor stars in the small box marked as red dots. The middle panel shows the box of [Fe/H] and [Mg/Fe] we have imposed in our selection. The right panel shows the $\tau-T_{\rm eff}$ relation for these old Mg-poor stars, while the white dots represent the comparison set, which is composed of stars in the same [Fe/H] and [Mg/Fe] box but with $T_{\rm eff}~\textgreater$ 5400 K. The background color-coded map indicates the full stellar sample in our catalog. }
\label{fig:a1}
\end{figure*}

\bibliography{sample631}{}
\bibliographystyle{aasjournal}



\end{document}